\documentclass[twocolumn]{aastex631}
\usepackage{xspace}
\usepackage{subfigure}
\usepackage{booktabs}
\usepackage{longtable}
\usepackage{multirow}
\usepackage{rotating}
\usepackage{tabularx}
\usepackage{appendix}
\usepackage{amsmath}
\usepackage{xcolor}
\usepackage{placeins}

\newcommand{\ci}{[C{\scriptsize I}]\xspace}
\newcommand{\cii}{[C{\scriptsize II}]\xspace}
\newcommand{\nii}{[N{\scriptsize II}]\xspace}
\newcommand{\oi}{[O{\scriptsize I}]\xspace}
\newcommand{\oiii}{[O{\scriptsize III}]\xspace}

\newcommand{\arc}{\ensuremath{''}\xspace}

\newcommand{\coseven}{\mbox{CO(7--6)}\xspace}
\newcommand{\citwo}{\mbox{[C{\sc i}](2--1)}\xspace}
\newcommand{\cione}{\mbox{[C{\sc i}](1--0)}\xspace}

\begin{document}

\title{A census of star-formation and gas mass tracers in two lensed $z \sim 4$ dusty star-forming galaxies}

\author[0000-0002-0786-7307]{David Vizgan}
\affiliation{Department of Astronomy, University of Illinois, 1002 West Green Street, Urbana, IL, 61801, USA}

\author[0000-0001-7192-3871]{Joaquin~D. Vieira}
\affiliation{Department of Astronomy, University of Illinois, 1002 West Green Street, Urbana, IL, 61801, USA}
\affiliation{Department of Physics, University of Illinois, 1110 West Green Street, Urbana, IL, 61801, USA}
\affiliation{Center for AstroPhysical Surveys, National Center for Supercomputing Applications, 1205 West Clark Street, Urbana, IL, 61820, USA}

\author[0000-0003-3256-5615]{Justin~S.~Spilker}
\affiliation{Department of Physics and Astronomy and George P. and Cynthia Woods Mitchell Institute for Fundamental Physics and Astronomy, Texas A\&M University, 4242 TAMU, College Station, TX 77843-4242, USA}

\author[0000-0003-3195-5507]{Simon Birrer}
\affiliation{Department of Physics and Astronomy, Stony Brook University, Stony Brook, NY 11794, USA}

\author[0000-0002-4861-0081]{Nan Zhang}
\affiliation{Department of Physics, University of Illinois, 1110 West Green Street, Urbana, IL, 61801, USA}

\author[0000-0002-6290-3198]{Manuel Aravena}
\affiliation{Instituto de Estudios Astrof\'{\i}cos, Facultad de Ingenier\'{\i}a y Ciencias, Universidad Diego Portales, Av. Ej\'ercito 441, Santiago, Chile}
\affiliation{Millenium Nucleus for Galaxies (MINGAL)}

\author[0000-0002-0517-9842]{Melanie~A. Archipley}
\affiliation{Department of Astronomy and Astrophysics, University of Chicago, 5640 South Ellis Avenue, Chicago, IL, 60637, USA}
\affiliation{Kavli Institute for Cosmological Physics, University of Chicago, 5640 South Ellis Avenue, Chicago, IL, 60637, USA}

\author[0000-0002-3272-7568]{Jack~E. Birkin}
\affiliation{Department of Physics and Astronomy and George P. and Cynthia Woods Mitchell Institute for Fundamental Physics and Astronomy, Texas A\&M University, 4242 TAMU, College Station, TX 77843-4242, USA}

\author[0000-0002-4657-7679]{Jared Cathey}
\affiliation{Department of Astronomy, University of Florida, 211 Bryant Space Sciences Center, Gainesville, FL 32611, USA}

\author[0000-0002-8487-3153]{Scott C. Chapman}
\affiliation{Department of Physics and Atmospheric Science, Dalhousie University, Halifax, Halifax, NS B3H 3J5, Canada}

\author[0000-0002-9993-3796]{Veronica~J. Dike}
\affiliation{Department of Astronomy, University of Illinois, 1002 West Green Street, Urbana, IL, 61801, USA}

\author[0000-0002-0933-8601]{Anthony~H. Gonzalez}
\affiliation{Department of Astronomy, University of Florida, 211 Bryant Space Sciences Center, Gainesville, FL 32611, USA}

\author[0000-0002-2554-1837]{Thomas R. Greve}
\affiliation{Cosmic Dawn Center (DAWN), Technical University of Denmark, DTU Space, Elektrovej 327, 2800 Kgs Lyngby, Denmark}

\author[0000-0002-7472-7697]{Gayathri Gururajan}
\affiliation{Scuola Internazionale Superiore Studi Avanzati (SISSA), Physics Area, Via Bonomea 265, 34135, Trieste, Italy}
\affiliation{IFPU-Institute for Fundamental Physics of the Universe, Via Beirut 2, 34014, Trieste, Italy}

\author[0009-0008-8718-0644]{Ryley Hill}
\affiliation{Department of Physics and Astronomy, University of British Columbia, 6224 Agricultural Road, Vancouver, BC, Canada}

\author[0000-0001-6919-1237]{Matthew A. Malkan}
\affiliation{Department of Physics and Astronomy, University of California Los Angeles, Los Angeles, CA 90095-1547, USA}

\author[0000-0002-7064-4309]{Desika Narayanan}
\affiliation{Department of Astronomy, University of Florida, 211 Bryant Space Sciences Center, Gainesville, FL 32611, USA}
\affiliation{Cosmic Dawn Center at the Niels Bohr Institute, University of Copenhagen and DTU-Space, Technical University of Denmark}

\author[0000-0001-7946-557X]{Kedar~A. Phadke}
\affiliation{Department of Astronomy, University of Illinois, 1002 West Green Street, Urbana, IL, 61801, USA}
\affiliation{Center for AstroPhysical Surveys, National Center for Supercomputing Applications, 1205 West Clark Street, Urbana, IL, 61820, USA}
\affiliation{NSF-Simons AI Institute for the Sky (SkAI), 172 E. Chestnut St., Chicago, IL 60611, USA}

\author[0009-0003-6250-1396]{Vismaya Pillai}
\affiliation{Department of Physics and Astronomy, University of British Columbia, 6224 Agricultural Road, Vancouver, BC, Canada}

\author[0000-0001-8598-064X]{Ana C. Posses}
\affiliation{Department of Physics and Astronomy and George P. and Cynthia Woods Mitchell Institute for Fundamental Physics and Astronomy, Texas A\&M University, 4242 TAMU, College Station, TX 77843-4242, USA}

\author[0000-0001-6629-0379]{Manuel Solimano}
\affiliation{Centro de Astrobiolog\'ia (CAB), CSIC-INTA, Ctra. de Ajalvir km 4, Torrej\'on de Ardoz, E-28850, Madrid, Spain}

\author[0000-0002-3187-1648]{Nikolaus Sulzenauer}
\affiliation{Max-Planck-Institut für Radioastronomie, Auf dem Hügel 69, Bonn, D-53121, Germany}

\author[0000-0002-6922-469X]{Dazhi Zhou}
\affiliation{Department of Physics and Astronomy, University of British Columbia, 6224 Agricultural Road, Vancouver, BC, Canada}

\begin{abstract}

We present new and archival Atacama Large Millimeter/submillimeter Array (ALMA) observations of two strongly lensed dusty star-forming galaxies (DSFGs) selected from the South Pole Telescope survey, SPT0418-47 ($z = 4.225$) and SPT2147-50 ($z = 3.760$). We study the \cii, \coseven, \citwo, and, in SPT0418-47, $p$-H$_2$O emission, which along with the underlying continuum (rest-frame 160 $\mu$m and 380 $\mu$m) are routinely used as tracers of gas mass and/or star-formation rate (SFR). We perform a pixel-by-pixel analysis of both sources in the image plane to study the resolved Kennicutt-Schmidt relation, finding generally good agreement between the slopes of the SFR versus gas mass surface density using the different tracers. Using lens modeling methods, we find that the dust emission is more compact than the line emission in both sources, with \coseven and \citwo similar in extent and \cii the most extended, reminiscent of recent findings of extended \cii spatial distributions in galaxies at similar cosmic epochs. We develop the \citwo / \coseven flux density ratio as an observable proxy for gas depletion timescale ($\tau_{\rm dep}$), which can be applied to large samples of DSFGs, in lieu of more detailed inferences of this timescale which require analysis of observations at multiple wavelengths. Furthermore, the extended \cii emission in both sources, compared to the total continuum and line emission, suggests that \cii, used in recent years as a molecular gas mass and SFR tracer in high-$z$ galaxies, may not always be a suitable tracer of these physical quantities.

\end{abstract}


\section{Introduction} \label{sec:intro}

Since the late 1990s, wide-area sky surveys in the millimeter (mm) and sub-mm windows have uncovered a population of ultraluminous infrared galaxies (ULIRGs) that have proved important for studies of galaxy evolution \citep[e.g.][]{smail1997, hughes1998, barger1998, chapman2005, weiss2009}. This population of sources, known as ``submillimeter galaxies'' \citep[SMGs;][for a review]{blain2002}, or more generally as dusty star-forming galaxies \citep[DSFGs;][]{casey2014}, are claimed to be the dust-enshrouded progenitors of the most massive galaxies in the local universe and signposts for high-$z$ galaxy clusters \citep[][]{toft2014, miller2018}. They form stars at rates of hundreds to thousands of solar masses per year \citep{hodge2020} and their dust absorbs a significant proportion of all optical and UV radiation in the Universe, re-emitting it into the sub-mm \citep{casey2014}. These sources thus represent an important fraction of the cosmic star-formation history \citep{madau2014} from $z=1-4$ and beyond \citep{swinbank2014, zavala2021}. 

The effect of ``negative \textit{K}-correction" \citep[e.g.][]{franceschini1991, blain1993} is known to benefit the selection of DSFGs at high-redshift. However, a substantial population of high-redshift galaxies were predicted to be effected by strong gravitational lensing \citep{blain1997, negrello2007} in large surveys. These surveys were ultimately realized in recent decades by observatories such as the South Pole Telescope \citep[SPT;][]{carlstrom2011}, \textit{Herschel} \citep{Pilbratt2010}, the Atacama Cosmology Telescope \citep[ACT;][]{swetz2011}, and \textit{Planck} \citep{planck2014} which detected the first large samples of mm-selected lensed DSFGs \citep[e.g.][]{vieira2010, Negrello2010, marsden2014, canameras2015}. A SPT-selected sample \citep{vieira2013} of 81 gravitationally-lensed DSFGs was selected using flux density cutoffs $S_{\rm 1.4mm} > 20$ mJy, $S_{\rm 870\mu m} > 25$ mJy \citep{reuter2020} and is complete in spectroscopic redshifts \citep{weiss2013, reuter2020}. It is estimated that about 70\% of these SPT-selected DSFGs are strongly lensed \citep{spilker2016}, with the rest being protoclusters \citep[e.g.][]{miller2018, hill2020} or unlensed sources.

The first spectra of SPT-selected DSFGs obtained by \cite{weiss2013} and \cite{vieira2013} revealed a rich inventory of spectral lines from molecules such as carbon monoxide (CO), along with CO isotopologues ($^{13}$CO), neutral carbon ($\rm{C{\scriptsize I}}$), and even signatures of water (o-H$_2$O). Observations in the following decade revealed a diverse inventory of spectral lines within DSFGs, both by stacking the individual spectra of SPT-selected DSFGs \citep{spilker2014, reuter2023} and by searches for specific lines. Along with multi-\textit{J} transitions of CO molecules \citep[e.g.][]{aravena2016, gururajan2023} and signatures of water \citep{jarugula2019, jarugula2021, reuter2023}, these lines include the bright far-infrared fine-structure lines (FSLs) e.g. \cii at 158 $\mu$m, which are the cooling lines emerging from the dense, warm neutral interstellar medium \citep[][for review]{hollenbach1999, decarli2025} as it condenses and contracts. Surveys of \cii \citep{gullberg2015} and \nii \citep{cunningham2020} have been completed for many of the SPT-selected DSFGs, alongside observations of other FSLs (e.g. \oi 146 $\mu$m, \oiii 88 $\mu$m) in individual SPT-selected DSFGs \citep{marrone2018, debreuck2019, litke2022}.

The Atacama Large Millimeter/submillimeter Array (ALMA) characterizes the dusty nature of extreme high-redshift galaxies in ways that are impossible via the optical and near-infrared windows \citep[accessed by e.g. the James Webb Space Telescope (\textit{JWST});][]{gardner2006}. Furthermore, many of the lines emitted within the mm and sub-mm window, which are unencumbered by dust attenuation (e.g. FSLs such as \cii), are utilized by astronomers across a wide range of subfields as tracers of gas mass and star-formation rate (SFR) within galaxies. For instance, the CO(1-0) molecule is used as a tracer of molecular hydrogen (H$_2$) in galaxies across a wide range of physical scales \citep{narayanan2012, bolatto2013}, which has been applied to high-redshift galaxies, including SMGs \citep{aravena2016, birkin2021, stanley2023}. Mid-$J$ CO lines are often extrapolated down towards the $J=1$ line in order to estimate gas masses, thanks to extensive work on the CO spectral line energy distribution (SLED) in normal galaxies, high-redshift galaxies, and quasars alike \citep{bothwell2013, spilker2014, yang2017, boogaard2020, birkin2021, sulzenauer2021, reuter2023, molyneux2025}, in the absence of shocks or AGN. High-$J$ CO lines have shown promise as a tracer of SFR in high-redshift galaxies via a strong correlation with infrared (IR) luminosities \citep{liu2015, lu2015, yang2017, jarugula2019}, motivated and calibrated in part by \textit{Herschel} observations of local galaxies \citep[e.g.][]{rosenberg2015, kamenetzky2016}. 

Alternative tracers are also used to probe the gas mass reservoirs in galaxies, the most prominent being neutral carbon (\ci), which has been shown to reliably trace the contents of the cold molecular interstellar medium (ISM) up to high redshifts \citep{gerin2000, papadopoulos2004, walter2011, bothwell2017, valentino2018, gururajan2023}. Recent work by \cite{Dunne2022} and \cite{gururajan2023} attempted to cross-calibrate gas mass measurements derived from CO, \ci, along with dust and, in the latter work, \cii luminosity. \cii emission, in recent years, has also been studied observationally and theoretically as a tracer of molecular gas in galaxies \citep{zanella2018, madden2020, vizgan22a, gurman2024, khatri2024, casavecchia2025}. It is otherwise used extensively as a tracer of SFR in galaxies \citep[e.g.][]{delooze2014}, and particularly in high-redshift galaxies \citep{vallini2015, lagache2018}.

\coseven and \citwo emission can be detected by ALMA in high-redshift galaxies, typically within the same spectral window, due to their close rest frequencies of 806.65180 GHz and 809.34197 GHz, respectively, with the water line $p-$H$_2$O$(2_{1,1}-2_{0,2}$) at 752.0331 GHz, nearby enough to observe in another spectral window. As such, these lines have been the target of many high-redshift galaxy observations \citep[e.g.][]{andreani2018, boogaard2020, valentino2018, valentino2020a, jarugula2021, gururajan2022, gururajan2023} and in quasars \citep{decarli2022, scholtz2023}. When considering \ci as a tracer of the cold molecular gas \citep[e.g.][]{papadopoulos2004} alongside \coseven as a tracer of star-formation rate (SFR), it can be hypothesized that the ratio of these two lines is an effective proxy for the gas depletion timescale $\tau_{\rm dep}$, defined as $M_{\rm gas}$ divided by SFR, and because it is derived from a line ratio, this proxy would be essentially independent of magnification due to gravitational lensing. Molecular gas masses estimated via the \citwo line require a third observable, the \cione line ($\nu_{\rm rest} = 492.161$ GHz), as the ratio of these lines yields a measurement of the gas excitation temperature \citep{stutzki1997}, from which a total mass of carbon can be estimated \citep{weiss2003}, after applying a carbon-to-hydrogen abundance factor $X_{\rm \ci}$. 

The TEMPLATES Early Release Science (ERS) Program \citep[for an overview see][]{rigby2025} observed two SPT-selected DSFGs with \textit{JWST}, SPT0418-47 and SPT2147-50, alongside two lensed galaxies from the Sloan Digital Sky Survey (SDSS) Giant Arcs Survey \citep[SGAS; e.g.][]{sharon2020}, to study and resolve star-formation within a wide variety of galaxy environments, magnified by the phenomenon of gravitational lensing. As part of this project, ALMA observations of \coseven and \citwo in the two SPT sources were successfully taken to complement the then-incoming \textit{JWST} observations. Early works from TEMPLATES \textit{JWST} observations have already yielded important science results, from the farthest ever detection of polycyclic aromatic hydrocarbons (PAHs) in the early Universe \citep{spilker2023} to the first resolved maps of metallicity in these two DSFGs \citep{birkin2023}, and even unambiguous signatures of a merger in SPT0418-47 \citep{cathey2024}. 

The motivation fueling this work is to compare several putative gas and SFR tracers, taking advantage of the boosted spatial resolution from magnification via gravitational lensing, to inform the use of these observables as physical tracers in high-redshift galaxies. In Section \ref{sec:obs} we detail the new and archival ALMA observations of SPT0418-47 and SPT2147-50, calibrate our observables into gas masses and SFRs, and described the gravitational lens models developed for this work. In Section \ref{sec:res} we investigate the resolved Kennicutt-Schmidt relation \citep{kennicutt1998} in the image plane, and study the radial profiles of gas and dust emission in the source plane. We discuss the implications of our image and source plane analyses in Section \ref{sec:disc} and conclude the paper in Section \ref{sec:conc}. Throughout this work we assume the cosmology realization from the Planck 2018 results \citep{planck2020} of flat $\Lambda$CDM, $H_0 = 67.7$ km s$^{-1}$ Mpc$^{-1}$, $\Omega_0$ = 0.31, and $T_{\rm CMB} = 2.725 {\rm \  K}$.

\section{Observations and Data Analysis} \label{sec:obs}

\begin{figure*}[ht!]
    \centering
    \includegraphics[width=\textwidth]{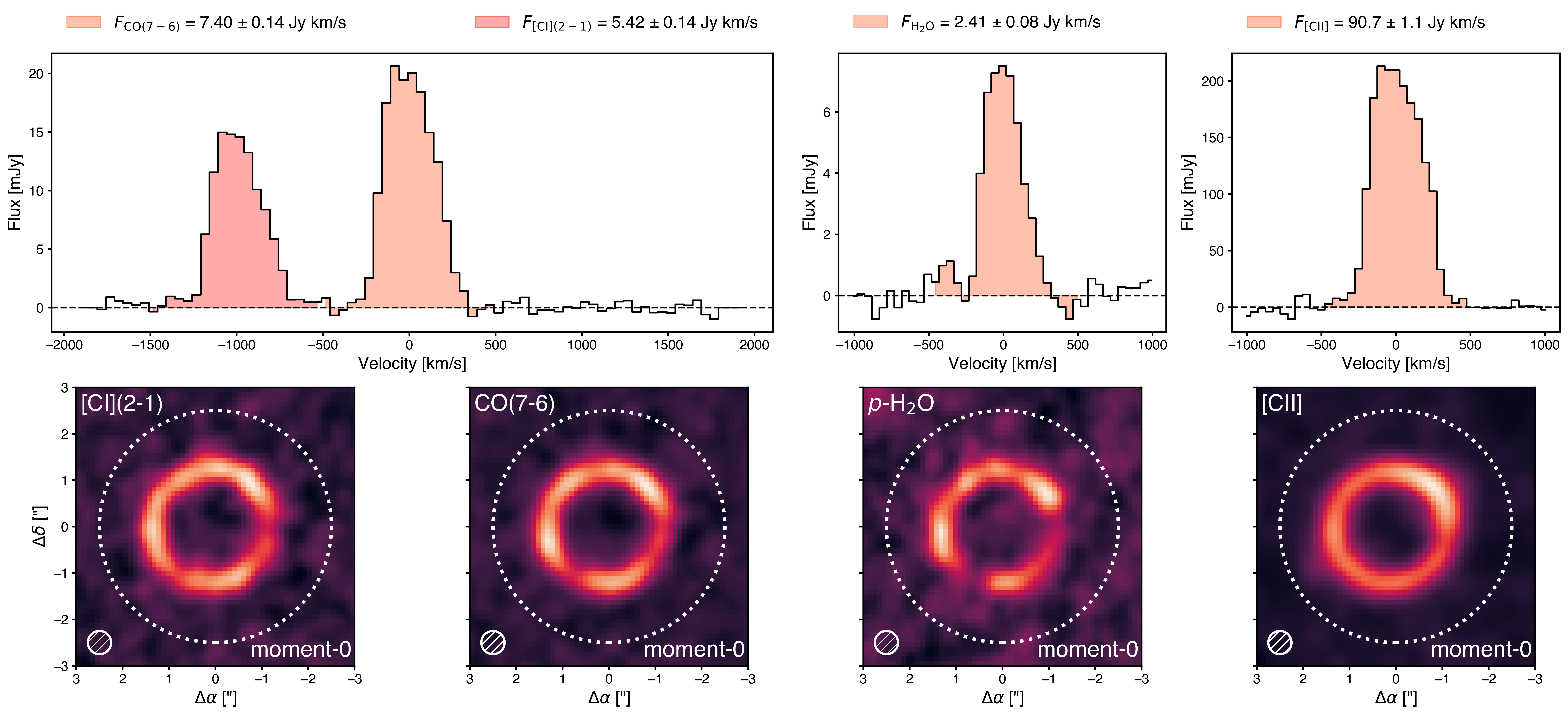}
    \caption{From left to right: \coseven, \citwo, $p$-H$_2$O, and \cii emission lines (top) with associated moment-0 maps (bottom). The spectra have a resolution of 50 km s$^{-1}$ and are generated from data cubes of SPT0418-47 within a circular aperture whose radius is 2.5\arc (as shown by the white circle). The cubes are re-imaged via \texttt{CASA} with a fixed circular restoring beam size of 0.5\arc and a pixel scale of 0.1\arc. All four lines are robustly detected and have similar line widths. Moment-0 maps are generated by integrating the channels highlighted in the spectra above, all encompassing 1000 km s$^{-1}$ in channel width.} 
    \label{fig:SPT0418-47_spectra}
\end{figure*}
\begin{figure*}
    \centering
    \includegraphics[width=\textwidth]{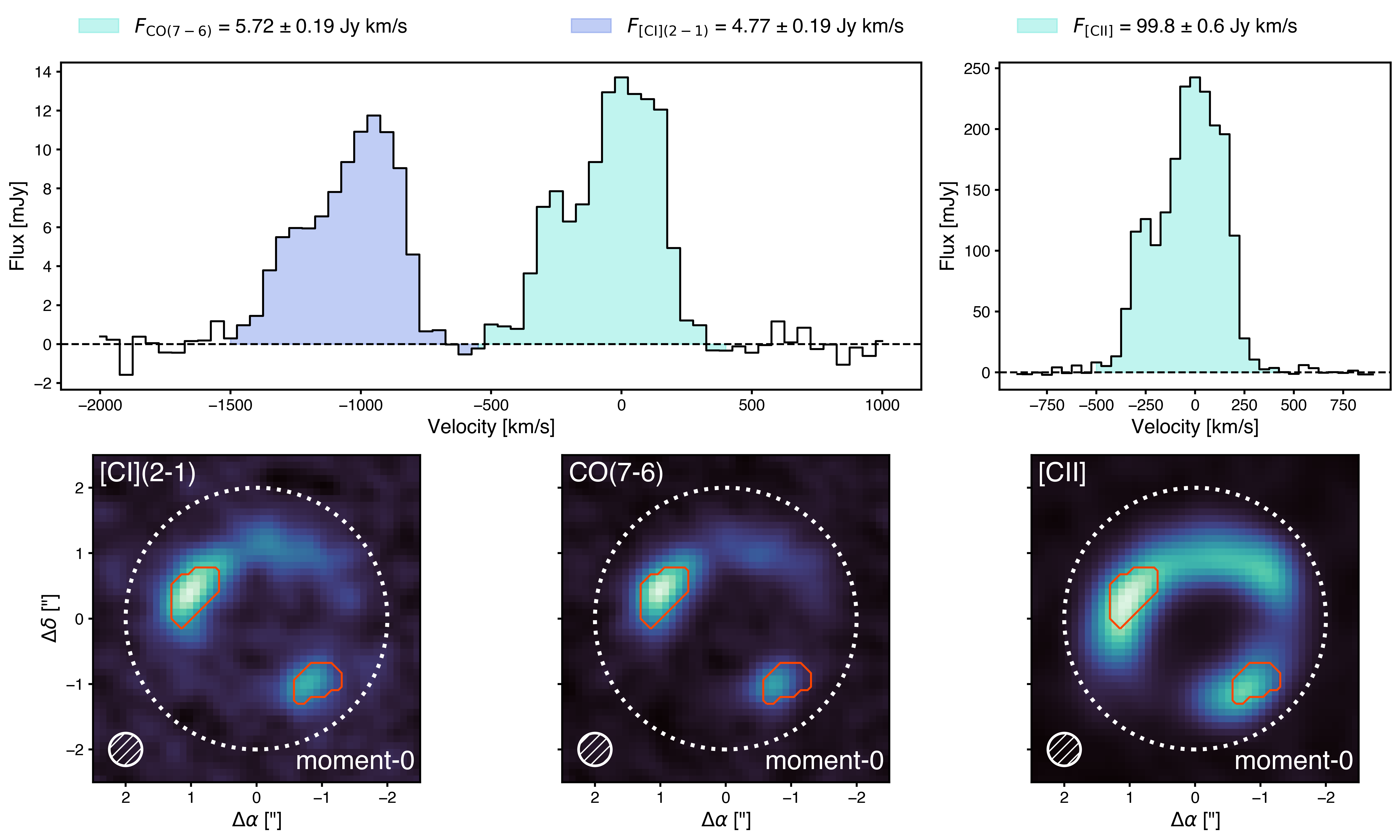}
    \caption{As in Figure 1, but for SPT2147-50. The aperture radius here, shown by the white circle, is 2\arc. We indicate, via the red contours, the location of the tentative AGN in \cite{birkin2023}. These pixels are removed for the fits in Figure \ref{fig:ks_law}.}
    \label{fig:SPT2147-50_spectra}
\end{figure*}

The observations presented in this work were taken by the ALMA 12-m array, located on the Chajnantor Plateau in Chile. In addition to new ALMA data of SPT0418-47 (2021.1.00252.S; PI: J. Vieira), taken as part of the TEMPLATES \textit{JWST} ERS program (along with 2018.1.01060.S; PI: Vieira), we utilize a combination of archival observations of \cii for this work (2016.1.01499.S; PI: Litke, 2016.1.01374.S; PI: Hezaveh, 2019.1.00471.S; PI: Spilker). 

For SPT0418-47 ($z = 4.2246 \pm 0.0004$, \cite{reuter2020}; $\alpha =$ 04:18:39.67, $\delta= $ $-$47:51:51.80), the \coseven, \citwo, and $p$-H$_2$O(2$_{1,1}$-2$_{0,2}$) lines were observed using the ALMA Band 4 Receiver \citep{asayama2014} from December 7 to December 9, 2021, across five execution blocks for a total of 5 hours and 20 mins. The correlator was configured to observe \coseven and \citwo in the upper sideband and H$_2$O in the lower sideband, with channel widths of 15.625 MHz and a total bandwidth of 2000 MHz per spectral window. The antennas spanned baselines (note: all baseline spans represent 5th and 80th percentiles) of 78 to 651 m, providing an approximate synthesized beam size of 0.47\arc $\times$ 0.44\arc via natural weighting, with the precipitable water vapor (PWV) ranging from 2.8--3.7 mm. The maximum recoverable scale of these observations is 6.05\arc. The \cii line was observed using the Band 7 receiver \citep{mahieu2012} from July 5, 2017 to November 6, 2017, across six execution blocks (two ALMA projects) for a total of 9 hours and 37 minutes, with the PWV varying between 0.2 and 1.5 mm.  The correlator was configured to observe \cii in the upper sideband with a channel width of 7.8125 MHz and total bandwidth of 1875 MHz per spectral window. In project 2016.1.01499.S, the array contained baselines spanning from 107--751 m, whereas in project 2016.1.01374.S the array contained baselines spanning from 101--635 m and 354--3783 m in two separate sets of observations. The maximum recoverable scale in these observations is 1.59\arc for the former project, and between 0.66\arc and 1.82\arc in the other project.

For SPT2147-50 ($z = 3.7604 \pm 0.0002$, \cite{reuter2020}; $\alpha =$ 21:47:19.01, $\delta =$ $-$50:35:54.35), the \coseven and \citwo lines were observed using the Band 5 receiver \citep{belitsky2018} from November 19 to 27, 2018, across three execution blocks for a total of 3 hours and 40 minutes. The correlator was configured to observe \coseven and \citwo in the upper sideband, with channel widths of 15.625 MHz and a total bandwidth of 2000 MHz per spectral window. The antennas spanned baselines of 54 and 489 m, providing an approximate synthesized beam of 0.62\arc $\times$ 0.48\arc via natural weighting, with the PWV ranging from 0.7--2.0 mm. The maximum recoverable scale for these observations is 7.04\arc. The \cii line was observed using the Band 8 receiver \citep{sekimoto2008} on May 4, 2019 and from May 24 to June 8, 2021, across four execution blocks (two ALMA projects) for a total of 7 hours and 27 minutes, with the PWV varying between 0.6 to 0.8 mm. The correlator was configured to observe \cii in the upper sideband with a channel width of 7.8125 MHz and total bandwidth of 1875 MHz per spectral window. In project 2019.1.00471.S, the array contained baselines spanning from 80--1056 m. The native resolution of both galaxies' \cii observations is $\approx$ 0.2\arc. For these two projects, the maximum recoverable scale ranges from 1.96\arc to 3.54\arc.

For all of the observations described above, nearby quasars were used to calibrate the observations for atmosphere, bandpass, phase, and gain. We use the software package \texttt{CASA} \citep[][]{mcmullin2007} to generate all data products for this work, utilizing the task \texttt{tclean} to create naturally weighted ``dirty'' and ``clean'' images of line emission. The phase center of project ID 2016.1.01374.S, targeting \cii emission in SPT0418-47, differed from project ID 2016.1.01499.S by 0.4\arc in right ascension and 2.6\arc in declination; in order to combine both visibility sets, this offset was corrected by using the mosaic gridder in the \texttt{CASA} task \texttt{tclean}. In Figure \ref{fig:SPT0418-47_spectra} and Figure \ref{fig:SPT2147-50_spectra} we present the \coseven, \citwo, \cii, and, for SPT0418-47, H$_2$O line emission. The images were created using the \texttt{CASA} task \texttt{tclean}, with a channel width of 50 km s$^{-1}$ for each respective datacube. 

We generate images with identical spatial resolution in order to enable a direct image plane (i.e. pixel-by-pixel) comparison. As such, we apply $uv$-tapers to each different band, varying from 0.2\arc to 0.5\arc, until we can achieve a circular restoring beam with a full-width at half-maximum (FWHM) of 0.5\arc. This resolution is chosen because it is close to the maximum achievable spatial resolution across both the \cii data and the lower resolution \coseven and \citwo datasets, with the primary goal to have beammatched data products for accurate image plane comparison. For SPT0418-47, the \coseven, \citwo, and $p$-H$_2$O data shown in Figure \ref{fig:SPT0418-47_spectra} were created using a 0.2\arc $uv$-taper and natural weighting, and the \cii data were created using Briggs weighting (robust $=$ 2.0) and a 0.5\arc $uv$-taper. In SPT2147-50, the \coseven and \citwo data shown in Figure \ref{fig:SPT2147-50_spectra} were created using Briggs weighting (robust $=$ 1.0) without a $uv$-taper, and the \cii data were created using Briggs weighting (robust $=$ 2.0) with a 0.4\arc $uv$-taper. The spectra shown in the top portions of Figures \ref{fig:SPT0418-47_spectra} and \ref{fig:SPT2147-50_spectra} are taken from all pixels within 2.5\arc and 2.0\arc apertures, centered at the source positions of SPT0418-47 and SPT2147-50, respectively. To calculate the total line fluxes, we extract a spectra from the datacubes within the aperture and sum up all channels which encompass the respective spectral line. We estimate the error on the lines via the standard deviation of flux in the line-free regions, which is propagated after accounting for channel summation.

In Figures \ref{fig:SPT0418-47_spectra} and \ref{fig:SPT2147-50_spectra}, the moment-0 maps are generated across 1000 km s$^{-1}$ (20 channels) for each detected line. While the spectral widths vary slightly from line to line in reality, we make this choice to ensure that we are encompassing the same physical regions of line emission across all observables. These maps were created via the \texttt{spectral-cube} Python package \citep{ginsburg2019}. We note here that \cite{birkin2023} showed evidence for an AGN in SPT2147-50 by studying \nii / H$\alpha$ maps in the galaxy, as observed by \textit{JWST}, finding elevated values of this ratio ($> 0.8$) along with broad H$\alpha$ emission in the source. We have highlighted the location of this AGN in Figure 2. When we perform our analysis of the spatially resolved Kennicutt-Schmidt relation in Section \ref{sec:res}, we do not include the pixels lying within the AGN region, in order to avoid AGN contamination.

\begin{deluxetable*}{lCCCCCCC}
\tablecaption{ALMA-derived line fluxes, FWHMs, and luminosities for SPT0418-47 and SPT2147-50.}
\label{tab:fluxes}
\tabletypesize{\small}
\tablehead{
\colhead{} &
\multicolumn{3}{c}{SPT0418-47 ($z=4.2246$)} &&
\multicolumn{3}{c}{SPT2147-50 ($z=3.7604$)} \\
\cline{2-4} \cline{6-8}
\colhead{Line} &
\colhead{$F_{\rm line}$ (Jy km s$^{-1}$)} &
\colhead{FWHM (km s$^{-1}$)} &
\colhead{$L_{\rm line}$ ($L_\odot$)} &&
\colhead{$F_{\rm line}$ (Jy km s$^{-1}$)} &
\colhead{FWHM (km s$^{-1}$)} &
\colhead{$L_{\rm line}$ ($L_\odot$)}
}
\startdata
CO(7–6) & $7.40 \pm 0.14$ & $296 \pm 4$ & $(1.82 \pm 0.03)\times10^9$ && $5.72 \pm 0.19$ & $385 \pm 5$ & $(1.17 \pm 0.04)\times10^9$ \\
\citwo & $5.42 \pm 0.14$ & $325 \pm 4$ & $(1.34 \pm 0.03)\times10^9$ && $4.77 \pm 0.19$ & $400 \pm 7$ & $(0.98 \pm 0.04)\times10^9$ \\
$p-$H$_2$O $(2_{1,1}-2_{0,2})$ & $2.41 \pm 0.08$ & $287 \pm 7$ & $(5.54 \pm 0.18)\times10^8$ && \nodata & \nodata & \nodata \\
\cii & $90.7 \pm 1.1$ & $352 \pm 2$ & $(5.27 \pm 0.06)\times10^{10}$ && $99.8 \pm 0.6$ & $379.4 \pm 0.9$ & $(4.81 \pm 0.03)\times10^{10}$ \\
\enddata
\end{deluxetable*}

In Table \ref{tab:fluxes} we present the fluxes and luminosities of the line emission from both sources. As the line profiles in both sources are not perfectly characterized by Gaussian profiles, we follow methodology in \cite{bothwell2013} for obtaining the equivalent FWHM of a non-Gaussian line profile using a Monte Carlo method. Across 2000 iterations, we take the spectral line within a $(-500, 500)$ km/s region, add Gaussian random noise, sampled from the error on the integrated flux, and calculate the second moment of this spectrum. From this sample of moments, we then take the mean and standard deviation, and multiply by a factor of 2.35 to obtain the resulting FWHM with uncertainty shown in Table \ref{tab:fluxes}. 

Flux values for the line emission in both sources were previously measured and published in \cite{gullberg2015, gururajan2022, gururajan2023} via ALMA, Atacama Pathfinder Experiment (APEX), and Atacama Compact Array (ACA) observations. Our line measurements in SPT2147-50 and SPT0418-47 are in agreement with the published data in \cite{gururajan2022, gururajan2023}, having lower uncertainties. In SPT0418-47 we determine a \cii luminosity that is $\approx 40\%$ lower than published \cii observations via single-dish APEX observations \citep[127 $\pm$ 10 Jy km s$^{-1}$;][]{gullberg2015}, and our high-resolution ALMA observations reveal, compared to \cite{gururajan2023}, $\approx 50\%$ brighter \citwo emission (5.42 $\pm$ 0.14 Jy km s$^{-1}$ vs. 3.57 $\pm$ 0.68 Jy km s$^{-1}$, respectively). The fainter \cii is easily explained by the much larger beam sizes of the APEX data \citep[13.5 to 22.0\arc in][]{gullberg2015} compared to our work. The brighter \ci, on the other hand, can be attributed in part to the fact that we achieve a much higher signal-to-noise ratio than \cite{gururajan2023} (38.7$\sigma$ vs. 5.25$\sigma$).

\subsection{Deriving physical quantities in the image plane}

For star-formation tracers, we first obtain pixel-by-pixel luminosities by converting pixel values in the moment-0 maps, with units of Jy km s$^{-1}$, into units of $L_{\rm \odot}$ via Equation 1 from \cite{solomon1997}:
\begin{equation}\label{eq:SolomonEq1}
    L_{\rm line} \ [{\rm L_{\odot}}]= 1.04 \times 10^{-3} S_{\rm line} \Delta v \ \nu_{\rm rest} {(1 + z)^{-1}} {D_{\rm L}}^2
\end{equation}
where $S_{\rm line} \Delta v$ represents the flux density of the line in Jy km s$^{-1}$, $\nu_{\rm rest}$ is the rest-frequency of a spectral line, $z$ is the redshift, and $D_{\rm L}$ is the luminosity distance. We utilize the following equations to convert from observed ALMA data products into SFRs. Following Equation 1 from \cite{lu2015},
\begin{equation}
    {\rm SFR} \ {\rm [M_{\odot} \ yr^{-1}]} = 1.31 \times 10^{-5}\ L_{\rm \scriptsize \coseven} \ {\rm [L_{\odot}]}
\end{equation}
converts \coseven luminosity to SFR; Equation 7 from \cite{jarugula2021},
\begin{equation}
    {\rm SFR} \ {\rm [M_{\odot} \ yr^{-1}]} = 2.07 \times 10^{-5}\ L_{\rm H_2O} \ {\rm [L_{\odot}]}
\end{equation}
converts H$_2$O luminosity to SFR; lastly, Equation 17 from \cite{delooze2014},
\begin{equation}
    \log_{10} {\rm SFR} = -8.52 + 1.18 \log_{10} L_{\rm \cii}
\end{equation}
converts \cii luminosity to SFR, with SFR and $L_{\rm \cii}$ measured in solar masses per year and solar luminosities, respectively. 

For gas mass tracers, we similarly utilize common conversion factors and equations from the literature. The general way to use CO molecules as gas tracers is by first multiplying CO luminosities (usually in units of K km s$^{-1}$ pc$^{2}$) with a conversion factor \citep[$X_{\rm CO}$ or $\alpha_{\rm CO}$; see][]{narayanan2012, bolatto2013}. We obtain these alternative units of luminosity via Equation 3 from \cite{solomon1997}:
\begin{equation}
    L'_{\rm line} \ {[\rm K \ km \ s^{-1} \ pc^2]} = 3.25 \times 10^7 \ S_{\rm line}\Delta v \ (\nu_{\rm obs})^{-2} {D_L}^2 (1 + z)^{-3}
\end{equation}
where $\nu_{\rm obs}$ is now the \textit{observed} frequency of a given spectral line. Using the modeled CO SLED of DSFGs from \cite{bothwell2013}, we divide our \coseven luminosities by a ratio $R_{\rm 71} =$ $L'_{\rm \scriptsize \coseven}/L'_{\rm CO(1-0)}=$ 0.18 \citep[see also][]{harrington2021}, and then multiply by $\alpha_{\rm CO} = 3.2$ $M_{\rm \odot}$ / (K km s$^{-1}$ pc$^2$), taken from \cite{Dunne2022} \citep[following work from][]{bothwell2017}. We note here that $\alpha_{\rm CO} = 0.8 \ M_{\rm \odot}$ / (K km s$^{-1}$ pc$^2$) is the standard value used for ULIRGs in the literature \citep[e.g.][]{downes1998}, and there exists a longstanding debate in the literature on the application of different $\alpha_{\rm CO}$ values to different types of galaxies. We also note that this value of $\alpha_{\rm CO}$ includes a factor of 1.36 to account for helium and other elements in molecular gas. We adopt a similar approach to convert \cii luminosity into molecular gas, albeit by first converting our moment-0 map of \cii into units of solar luminosities before applying a standard conversion factor of 31 $\pm$ 6 M$_{\rm \odot} / $ L$_{\rm \odot}$ \citep{zanella2018}.

To convert \citwo into molecular gas mass, we follow Equation 3 from \cite{weiss2003},
\begin{equation}
    M_{\rm \ci} \ {\rm [M_{\odot}]}= 4.566 \times 10^{-4} \ Q(T_{\rm ex}) \ \frac{1}{5} e^{62.5 / T_{\rm ex}} L'_{\rm \scriptsize \citwo}
\end{equation}
with \citwo luminosity in units of K km s$^{-1}$ pc$^2$. Here, $Q$ represents the partition function of \ci,
\begin{equation}
    Q(T_{\rm ex}) = 1 + 3e^{-23.6  / T_{\rm ex}} + 5e^{-62.5 / T_{\rm ex}}
\end{equation}
with $T_{\rm ex}$ representing the excitation temperature in units of Kelvin. This temperature, in turn, can be approximated \citep{stutzki1997, schneider2003} as
\begin{equation}
    T_{\rm ex} \ [{\rm K}] = \frac{38.8}{\ln (2.11 / R_{\rm \ci})}
\end{equation}
with $R_{\rm \ci}$ representing the line luminosity ($L'$) ratio of \citwo and \cione, assuming that both \ci lines are optically thin. While we do not have spatially resolved \cione, \cite{bothwell2017} presented integrated \cione luminosities for 13 DSFGs, including SPT0418-47 and SPT2147-50. We thus use these data to calculate a global ratio of \ci luminosities (units of K km s$^{-1}$ pc$^2$) in both sources, to then calculate an excitation temperature. 

Using the values in Table \ref{tab:fluxes}, we obtain values of $R_{\rm \ci} = 0.8 \pm 0.2$ and $0.9 \pm 0.3$ for SPT0418-47 and SPT2147-50, respectively, and from these ratios we derive excitation temperatures of $41\pm11$ K and $44\pm15$ K for SPT0418-47 and SPT2147-50, respectively. These values are in agreement with \cite{gururajan2023}, who determined excitation temperature values ranging from 17.7 to 64.2 K in a sample of SPT-selected SMGs from unresolved measurements of $R_{\rm \ci}$. The excitation temperatures here are lower than the dust temperatures estimated in \cite{reuter2020} for both sources (58 $\pm$ 11 K and 48 $\pm$ 9 K, respectively) but overlap within uncertainties. It is argued in \cite{gururajan2023}, via unresolved \ci observations of SPT SMGs, that the difference in these temperatures can be explained by \ci emission extending into cooler regions compared to the dust. This explanation is compelling when considering that, as shown later in this work, the modeled \ci emission indeed extends beyond the dust continuum for both SPT0418-47 and SPT2147-50. Finally, we assume a ratio of $X_{\rm \ci} = M_{\rm \ci} / M_{\rm H_2} \approx 10^{-3.9 \pm 0.1}$ as derived from CO(2-1) observations \citep{aravena2016} of SPT-selected DSFGs \citep{bothwell2017} by \cite{valentino2018}, and multiply by a factor of 1.36 to account for the presence of helium and heavier elements. We note that $X_{\rm \ci} =  10^{-3.9}$ is at the highest end of $X_{\rm \ci}$ values in the literature \citep{Dunne2022}.

\subsection{Gravitational lens models}

Having analyzed the spectra and the emission line maps of both SPT0418-47 and SPT2147-50, we now turn towards the source plane analysis for both sources. To complement this pixel-by-pixel analysis, we use the Python package \texttt{lenstronomy} \citep{birrer2018, birrer2021} to develop lens models of SPT0418-47 and SPT2147-50. We specifically utilize a modification to \texttt{lenstronomy}, as detailed in \citep{zhang2025} which enables the software to create lens models of images that are made from interferometric data (e.g. ALMA visibilities). Although there are adequate lens models for both sources existing in the literature \citep[e.g][and others]{spilker2016}, we remake lens models in this section to take advantage of our deep multi-band observations with ALMA. 

For lens modeling, we create ``dirty'' images from the combined visibility data, beam-matching the higher resolution \cii data to the native \coseven resolution, using all channels containing line emission combined via the multi-frequency synthesis mode of the \texttt{CASA} task \texttt{tclean}. We do this because \texttt{lenstronomy} can only model data generated from naturally weighted dirty images, and some of the products shown in Figures \ref{fig:SPT0418-47_profiles} and \ref{fig:SPT2147-50_profiles} were made with Briggs weighting and with \textit{uv}-tapering. In addition to all of the emission lines shown in Figures \ref{fig:SPT0418-47_spectra} and \ref{fig:SPT2147-50_spectra}, we generate maps of the associated continuum emission at 160 and 380 $\mu$m to study the dust vs. gas in both sources. For each band we model the emission from the source as a S\'ersic profile \citep{sersic1963}. For the foreground source, we do not model any light in the profile and assume a singular isothermal ellipsoid \citep[SIE;][]{kormann1994} for the foreground mass profile.

The modeled parameters in each source, as presented in Tables \ref{tab:parameters_0418} and \ref{tab:parameters_2147} in the Appendix, are the
\begin{itemize}
    \item Einstein radius $\theta_E$
    \item lens ellipticity parameters $e_1, e_2$
    \item lens position $x$, $y$ relative to measured right ascension and declination of system (see Section \ref{sec:obs})
    \item shear parameters $\gamma_1$, $\gamma_2$
    \item S\'ersic half-light radius $r_e$ for each band
    \item S\'ersic index $n$ for each band
    \item source ellipticity parameters $e_1$, $e_2$
    \item source position $x, y$ relative to measured right ascension and declination of system
\end{itemize}

The foreground lens modeling properties are shared across a joint fit to each band, while the source ellipticity parameters and central coordinate are shared across all bands of source emission, in order to fix a shared physical geometry for the source. For simplicity, we fix the center of external shear to the system's positional center (i.e. $\Delta x$, $\Delta y$ $=$ 0,0) during the fitting. We note that the fitting of external shear yields a modest, if not negligible, effect on both lensing systems (see Appendix). 

The models utilize Markov Chain Monte Carlo methods (MCMC) via \texttt{emcee} \citep{foremanmackey2013}, which is wrapped into \texttt{lenstronomy} to retrieve posterior distributions for each modeled parameter. We initialize a particle swarm optimization \citep[PSO;][]{kennedy1995} for 200 particles, after which 50 walkers run across the parameter space until most of the parameter chains reach 50 $\times$ the autocorrelation timescale $\tau$. The best-fit model is determined by $\chi^2$ minimization. Visualizations of the best-fit lens models are presented in the Appendix, with the posterior distributions and priors presented in Tables \ref{tab:parameters_0418} and \ref{tab:parameters_2147}. 

\section{Results}\label{sec:res}

\subsection{The spatially resolved Kennicutt-Schmidt relation}
\begin{figure*}[h!]
    \centering
    \includegraphics[width=\textwidth]{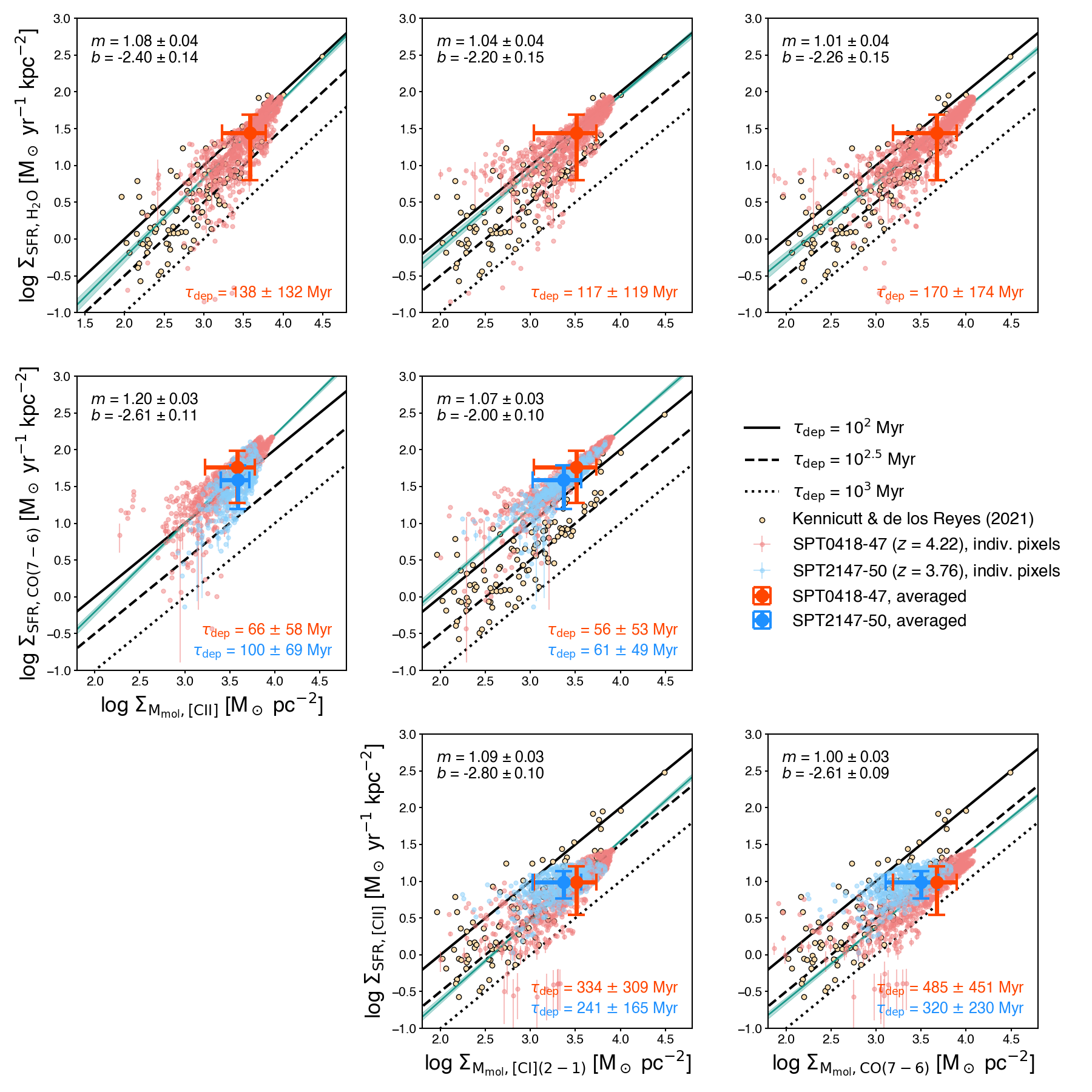}
    \caption{The spatially resolved Kennicutt-Schmidt law as derived from our sample of observed lines (Table \ref{tab:fluxes}). Each row shares the same $y$-axis. We find generally consistent linear slopes between log $\Sigma_{\rm SFR}$ and $\Sigma_{\rm H_2}$, regardless of which line is used to trace either quantity. The ivory-colored circles represent starbursts from \cite{kennicutt2021}. The red (SPT0418-47) and blue (SPT2147-50) points represent all pixels, selected from the maps of line emission, with the large points indicating globally integrated values. The green line demonstrates the best-fit line to the combined pixel-by-pixel data in order to compare with the canonical Kennicutt-Schmidt law. We lastly add straight lines that represent gas depletion timescales of 10$^{2, \ 2.5, \ 3}$ Myr. All pixels which spatially coincide with the likely AGN in SPT2147-50 \citep[][see Appendix]{birkin2023} have been omitted from this plot. }
    \label{fig:ks_law}
\end{figure*}

In Figure \ref{fig:ks_law} we plot the resolved Kennicutt-Schmidt relation -- via converted SFR surface densities ($\Sigma_{\rm SFR}$) and molecular gas mass densities ($\Sigma_{\rm H_2}$) -- in SPT0418-47 and SPT2147-50. To perform this analysis, we first mask all pixels where 380 $\mu$m continuum emission is $< 5\sigma$. Then, we calculate values of $\Sigma_{\rm SFR}$ and $\Sigma_{\rm H_2}$ on a pixel by pixel basis, using the calibration factors and equations listed above. We plot every single pixel in the emission line maps of both galaxies and plot the averaged value for each surface density, including the propagated uncertainty both from the conversion factors/equations and the spread in pixel values. We average the pixels for two reasons. First, some regions of emission which are spatially separate in the image plane arise from the same physical region in the source plane, which is being magnified along the lensing caustic. Additionally, our data products have a pixel size of 0.1\arc, which is lower than the data beam sizes of 0.5\arc; because of this, not all pixels are statistically independent, and averaging all of the values allows us to reduce autocorrelation between pixels. 

Each sub-panel in Figure \ref{fig:ks_law} represents different emission lines converted into gas mass or star-formation rate surface densities. For comparison, we have plotted a sample of local starburst galaxies taken from \cite{kennicutt2021} which demonstrate that fitting the canonical Kennicutt-Schmidt relation to starburst galaxies returns a power-law slope index $m = 0.98 \pm 0.07$ \citep{kennicutt2021}. To compare with their work, we perform line fitting to the pixels in both SPT0418-47 and SPT2147-50 in order to derive the characteristic slope $m$ of the Kennicutt-Schmidt relation. We normalize the fits onto the medians of the pixels before fitting in order to reduce degeneracy between the fitted slope and intercept. 

Since we have estimated both SFR and molecular gas masses in these sources, it is possible to estimate the gas depletion timescale ($\tau_{\rm dep}$) in both SPT0418-47 and SPT2147-50. The resulting value in both sources is important, as starbursts, in addition to high star-formation rates, are characterized by their low gas depletion times. Using low-$J$ CO lines, \cite{aravena2016} showed that the gas depletion timescale in a sample of lensed SPT-selected SMGs, including the sources in this work, is $< 200$ Myr, and \cite{yang2017} found a similar timescale ($\approx 20-100$ Myr) in \textit{Herschel}-selected lensed SMGs using mid-$J$ CO lines and \ci lines. From the individual pixels in our image plane analysis, we calculate gas depletion times for each possible combination of gas mass and SFR tracers. For all combinations of tracers, with the exception of \cii vs. \ci, we find gas depletion timescales $<$ 200 Myr, with yet another caveat that the gas depletion timescales have limited precision due to the multiple sources of error included for each calculation of $\tau_{\rm dep}$.

For the most part, our fitted relations (green line) show excellent agreement to the literature \citep[with $1.00 < m < 1.09$ in six of the seven calibrations, compared to $m \approx 1$ in][]{kennicutt2021}. \cii as both a gas mass and SFR tracer shows the largest deviations from the Kennicutt-Schmidt relation. As a gas mass tracer vs. \coseven as a SFR tracer, we find a characteristic Kennicutt-Schmidt slope $m \approx 1.20$. As a SFR tracer vs. both \citwo and \coseven, we obtain unrealistic gas depletion times (hundreds of Myr) for what are expected to be starburst galaxies. In the latter case, SFRs derived from \cite{delooze2014} are lower than those derived from H$_2$O and \coseven emission, indicating that this relation may not work for starbursts at spatially-resolved scales. As aforementioned, we have removed the AGN region in SPT2147-50 as speculated in \cite{birkin2023} in order to perform this analysis, as several of the equations used to estimate SFRs and gas masses should not be applicable to AGN; in the Appendix, we show that removing the AGN pixels from SPT2147-50 does not significantly affect the line fits.

\subsection{Radial profiles of dust and gas}\label{sec:sou}

\begin{figure*}
    \centering
    \includegraphics[width=\textwidth]{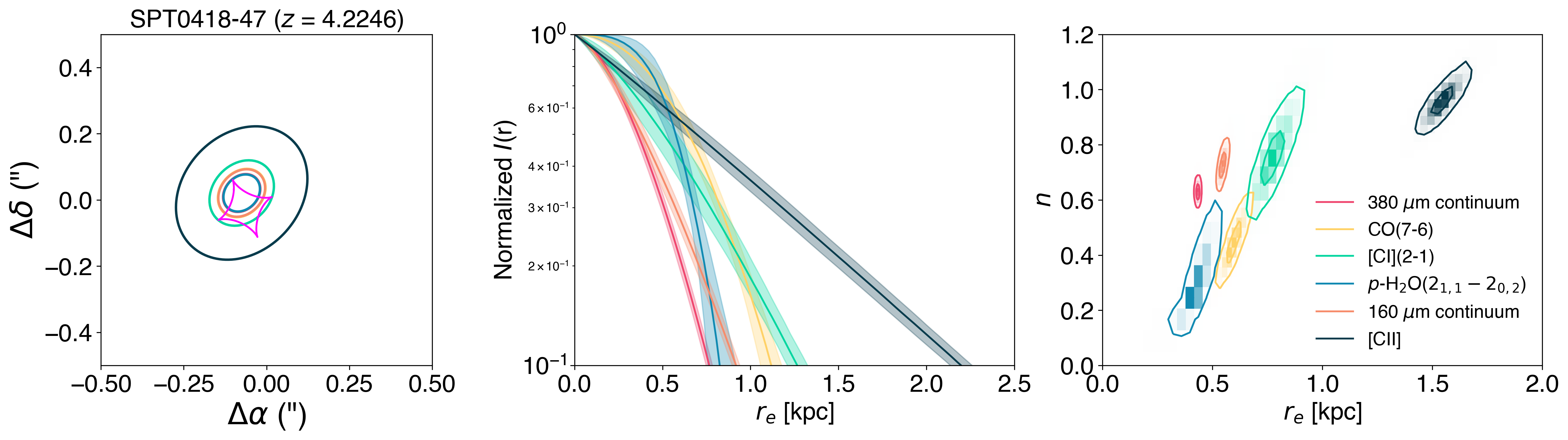}
    \caption{\textbf{Left:} Best-fit half-light ellipses derived from lens modeling (see Table \ref{tab:parameters_0418}) for each modeled band of dust and line emission in SPT0418-47, projected onto the sky; the best-fit lensing caustic is shown in magenta. \textbf{Middle:} Median best-fit S\`ersic profiles (solid lines) as a function of source plane radius in kpc. Uncertainties are determined via 1000 random pulls from the posterior distributions, as outputted by the MCMC fitting. \textbf{Right}: 2D histograms of the half-light radius vs. S\`ersic index for each profile, with 1-sigma and 2-sigma contours overlaid. Contours are drawn via kernel density estimation assuming a 2D Gaussian probability density.}
    \label{fig:SPT0418-47_profiles}
\end{figure*}

\begin{figure*}
    \centering
    \includegraphics[width=\textwidth]{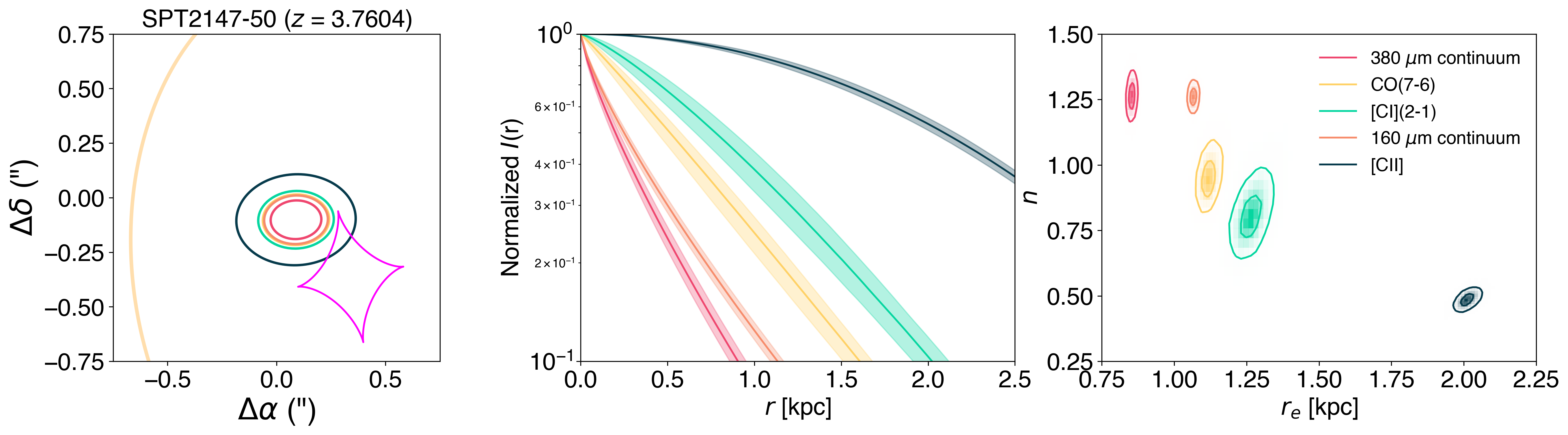}
    \caption{Same as in Figure \ref{fig:SPT0418-47_profiles}, but for SPT2147-50. The best-fit critical curve is shown in gold.}
    \label{fig:SPT2147-50_profiles}
\end{figure*}

In Figures \ref{fig:SPT0418-47_profiles} and \ref{fig:SPT2147-50_profiles} we plot the normalized S\'ersic profiles for each modeled band in SPT0418-47 and SPT2147-50, respectively. The uncertainties on the S\`ersic index and radius are estimated and plotted via bootstrapping 1000 random draws from the chains, and taking one standard deviation from those points. The profiles are normalized to the peak value of each bootstrapped profile. 

We can observe three important features about the internal structure of these two DSFGs. First, the \coseven emission in both sources shares the same spatial extent as the \citwo line. This is not necessarily surprising; if one assumes that \ci is tracing the cold molecular hydrogen, and that high-$J$ CO is tracing the active star-forming activity in galaxies, it is natural to expect both lines to arise from the same regions, as star-formation is fueled by reservoirs of molecular hydrogen. Second, the continuum emission at 160 $\mu$m and 380 $\mu$m share a similar spatial extent, and both are more compact than the emission lines. This is not surprising, as \cite{gururajan2022} found that \coseven and \citwo were more spatially extended than the dust continuum in 3 gravitationally lensed galaxies at $z\ge3$, including SPT2147-50. 

Lastly, we find in both galaxies that the \cii line emission is spatially extended compared to all of the other modeled bands, with an effective radius $r_e = 1.55^{+0.07}_{-0.06}$ kpc and $2.01^{+0.03}_{-0.02}$ kpc for SPT0418-47 and SPT2147-50, respectively. It is shown in the rightmost panel of Figures \ref{fig:SPT0418-47_profiles} and \ref{fig:SPT2147-50_profiles} that the \cii emission in both sources is far more extended than for all other profiles. We further discuss the extended \cii emission and how it may be interpreted in Section \ref{sec:disc}. To draw the contours shown in the right-most panels, we use the \texttt{corner} Python package \citep{corner}.

\cite{yang2017, yang2019} argue that mid-$J$ CO emission and low-$J$ H$_2$O emission are expected to be co-spatial due to similar linewidths, and can in turn be used as a gas-to-dust mass ratio. In SPT0418-47, however, we find that the water emission appears much more spatially truncated compared to the \coseven and \citwo line emission; while the water emission seems to roughly trace the dust emission, the low Sersic index indicates that the bulk of the water emission arises directly from the core of the galaxy, and although the water has a much more compact spatial extent than the CO emission in SPT0418-47, the best-fit lens models (see Figures. \ref{fig:SPT0418-47_profiles} and \ref{fig:SPT2147-50_profiles}) agree with the interpretation of \cite{yang2017, yang2019} that low-$J$ water lines are pumped up by FIR emission which in turn arises from the dust within the SMG environment. 

\section{Discussion} \label{sec:disc}

\subsection{\coseven, \citwo, and the gas depletion timescale in high-redshift galaxies}

Gas depletion timescales were derived for the SPT-selected DSFG sample in \cite{reuter2020}, with a median $\tau_{\rm dep} \approx 59$ Myr. The gas masses in their work were derived by assuming a dust-to-gas ratio of 100, rather than by calibrating spectral line observations to derive molecular gas masses. We have shown in Section \ref{sec:obs} that extensive work in the literature focuses on calibrating different line luminosities as tracers of the molecular gas reservoirs and the star formation rate in high-redshift, star-forming galaxies. SED-derived $\tau_{\rm dep}$ derivations require many observations that can cover most of the millimeter and submillimeter window, often requiring multiple observatories (e.g. \textit{Herschel}) aside from ALMA to constrain the IR luminosity and dust mass. 

In this section we aim to develop the \citwo to \coseven flux density ratio as a proxy of the gas depletion timescale in high-redshift galaxies. In our image plane analysis, we have shown that the use of \coseven as a SFR tracer in tandem with \citwo as a molecular gas mass tracer returns a roughly linear Kennicutt-Schmidt slope ($n = 1.07 \pm 0.03$) plus relatively low gas depletion timescales ($\tau_{\rm dep} \lesssim 100$ Myr), all in accordance with predictions for local starbursts. In our source plane analysis, the radial extent of \coseven and \citwo covers roughly the same region of the source compared to the more compact dust continuum and more extended \cii emission. The combination of these three factors makes the \citwo to \coseven ratio a  convenient proxy for the gas depletion timescale. 

There are some recent but notable counter-examples in the literature of \citwo and \coseven tracing different regions. \cite{huang2024} found differing morphologies of \coseven and \citwo in a starburst galaxy at $z = 2.57$, where a blob spatially offset to the south of the source features strong CO(7-6) emission and no \citwo emission, although a tidal tail in \citwo extends towards the blob. In a sample of 15 extremely red quasars at $ z \sim 2.3$, \cite{scholtz2023} found a wide range of spatial extents when comparing \coseven and \citwo emission, with direct evidence for matching \coseven and \citwo emission in only three sources. These examples serve as a counter-argument for using the \citwo / \coseven flux density ratio as a proxy for $\tau_{\rm dep}$, since in these sources, the measured lines do not always arise from the same physical region. They also highlight the importance of studying spatially resolved \coseven and \citwo in larger samples of high-redshift galaxies. 

\begin{figure*}
    \centering
    \includegraphics[width=\textwidth]{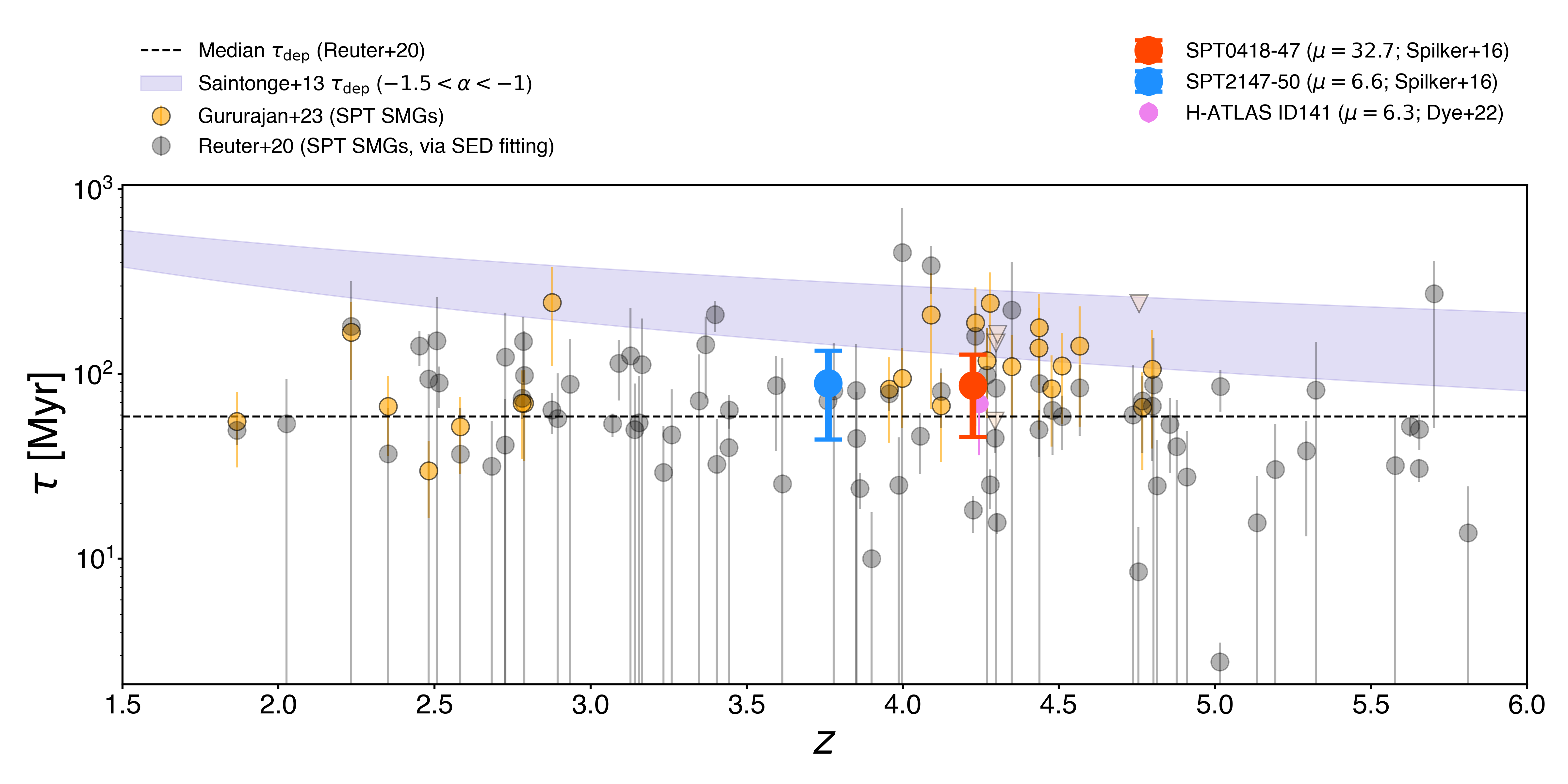}
    \caption{Gas depletion timescale $\tau_{\rm dep}$ as a function of redshift $z$. We include our work (SPT0418-47 in red, SPT2147-50 in blue), a sample of SPT-selected DSFGs \citep[][in orange]{gururajan2023}, and H-ATLAS ID141 \citep{dye2022}. We plot the model from \cite{saintonge2013} in lavender to show the predicted evolution of $\tau_{\rm dep}$ vs. cosmic time in main sequence galaxies. We have additionally include the IR-derived gas depletion times from \cite{reuter2020} in gray. The transparent triangles represent upper limits on line detections.}
    \label{fig:tdepvsz}
\end{figure*}

Converting \coseven to SFR is straightforward as the line is strongly correlated with FIR luminosity \citep[e.g.][and Figure \ref{fig:tdepvsz}]{liu2015}. Converting \citwo into a molecular gas mass estimate, however, is more challenging. With \cione in a different ALMA Band compared to \coseven and \citwo, it is not always possible to determine the ratio of both \ci lines in order to derive an excitation temperature. It is additionally a challenge to directly observe a carbon-to-molecular gas abundance ratio ($X_{\rm \ci} = M_{\rm \ci} / M_{\rm H_2}$) in high-redshift galaxies, and a value of $X_{\rm \ci}$ is typically applied from the literature to observations. With all of these caveats in mind, we combine equations from \cite{solomon1997, weiss2003, lu2015}, as reproduced in Section \ref{sec:obs} of this work, to derive an equation that can estimate gas depletion times in high-redshift galaxies via the \coseven and \citwo lines:
\begin{align}
    \tau_{\rm dep} \ [{\rm Myr}] &= \frac{2.80 \times 10^{-3}}{X_{\rm \ci}} \notag \\
    &\quad\times \left[ 
    0.66 \left(\frac{1}{R_{\rm \ci}} \right)^{1.61} + 1.27\left(\frac{1}{R_{\rm \ci}}\right) + 1 \right]  \notag \\
    &\quad\times \frac{F_{\rm \scriptsize \citwo}}{F_{\rm \scriptsize \coseven}}
\label{eq:DV-tdep}
\end{align}

\begin{figure}
    \centering
    \includegraphics[width=0.5\textwidth]{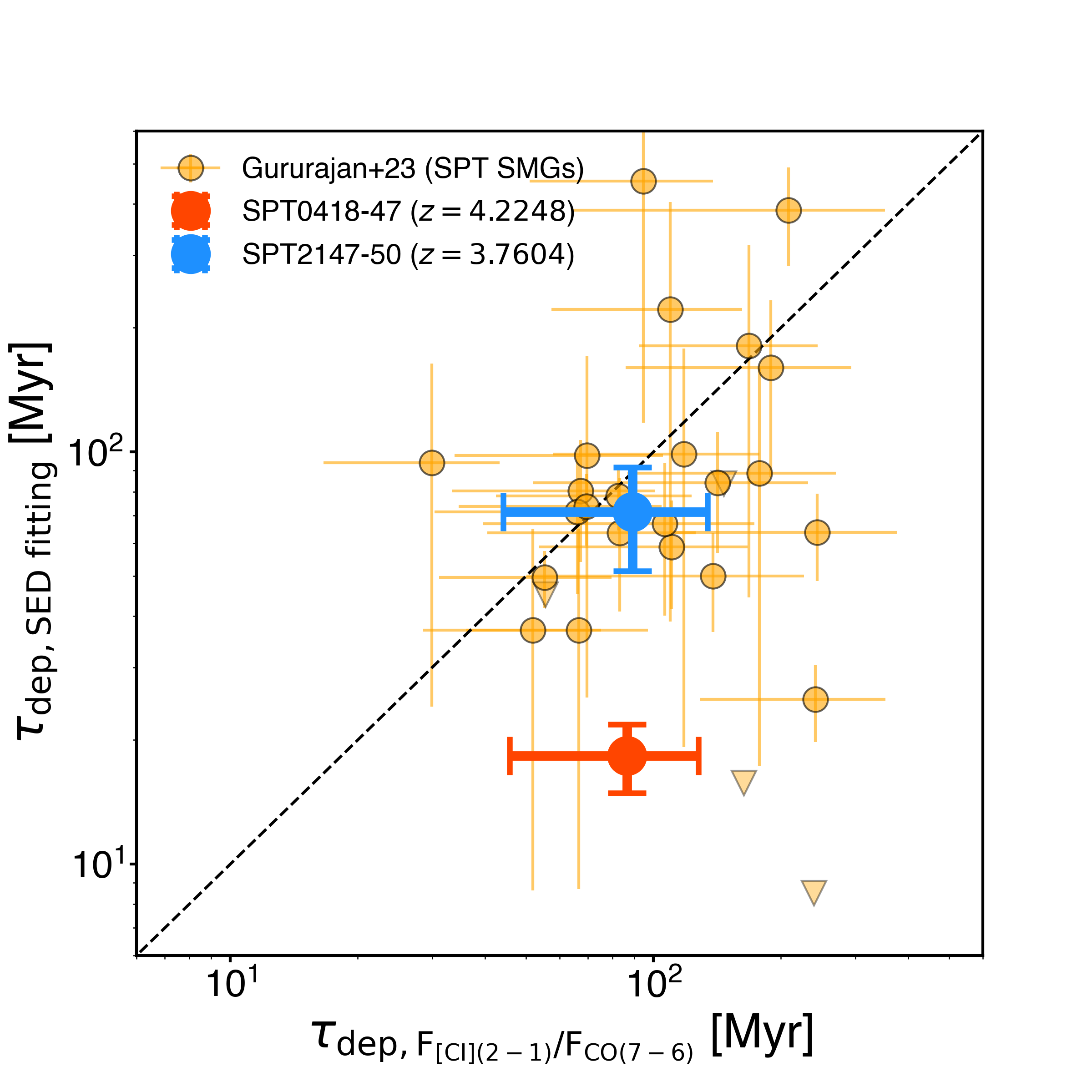}
    \caption{Gas depletion timescales as computed by SED fitting \citep{reuter2020} vs. the ratio of \citwo and \coseven flux densities, as taken from \citep{gururajan2023}. We also show a sub-sample of constituent galaxies of SPT2349-56 \citep{hill2022, hughes2025}. The TEMPLATES galaxies (shown in red and blue) studied in this paper have been removed from the unresolved \cite{gururajan2023} sample and replaced with our derived flux densities from resolved observations. Both methods are plagued with large uncertainties due to error on dust mass via SED fitting, and poor constraints on $X_{\rm \ci}$ and $R_{\rm \ci}$ in observed DSFGs respectively. Nonetheless, the derived gas depletion times agree with each other within uncertainties. The dashed black line represents a 1:1 ratio. Transparent triangles represent sources in which only upper limits of at least one line flux density are available.}
    \label{fig:comparison_tdep}
\end{figure}

In Figure \ref{fig:tdepvsz} we compare our derived gas depletion times from Equation \ref{eq:DV-tdep} across $z$ for the objects presented our work, the sample of SPT sources presented in \cite{gururajan2023}, and H-ATLAS ID141 \citep{dye2022}, a DSFG at a similar redshift to our sources. Generally, we find that the ratio of \citwo to \coseven flux densities alone are able to recover gas depletion timescales that are comparable to those estimated via SED-fitting to the FIR emission \citep{reuter2020}. A visualization of this comparison is shown in Figure \ref{fig:comparison_tdep}. In sources from \cite{gururajan2023} without a \cione or \citwo detection, we use a $R_{\rm \ci} = 0.69 \pm 0.05$, derived by converting the mean excitation temperature found in their work ($T_{\rm ex} =$ 34.5 $\pm$ 2.1 K) to a \ci line ratio. 

We can also compare our flux density-derived $\tau_{\rm dep}$ with \cite{aravena2016}, who calculated gas depletion times from gas masses via CO(2-1) observations, and SFRs via IR luminosities, assuming the initial mass function from \cite{chabrier2003}. In SPT0418-47, we find $\tau_{\rm dep} = 18 \pm 5$ Myr via \cite{reuter2020}, $\tau_{\rm dep} = 21 \pm 6$ Myr via \cite{aravena2016}, and $\tau_{\rm dep} = 87 \pm 41$ Myr in this work. In SPT2147-50, we find $\tau_{\rm dep} = 70 \pm 20$ Myr via \cite{reuter2020}, $\tau_{\rm dep} = 35 \pm 10$ Myr via \cite{aravena2016}, and $\tau_{\rm dep} = 89 \pm 45$ Myr in this work. 

The gas depletion timescales derived from Equation \ref{eq:DV-tdep} are approximately a factor of two larger than the FIR-derived and low-$J$ CO-derived $\tau_{\rm dep}$ in \cite{aravena2016, reuter2020}. We find that the primary driver of difference in our derived gas depletion timescales comes from the assumed $X_{\rm \ci}$ value which has been the subject of many other works \citep{bothwell2017, valentino2018, valentino2020a, Dunne2022}. and is indeed an exploration which is mostly beyond the scope of this work. We note, though, that the above calculation, along with the calculations used in Figures \ref{fig:tdepvsz} and \ref{fig:comparison_tdep} uses a $X_{\rm \ci} = 8.4 \pm 3.5 \times 10^{-5}$ from \cite{walter2011}, and if we assume $X_{\rm \ci} \approx 1.7 \times 10^{-4} \approx 10^{-3.8}$, a value in line with \cite{valentino2018} ($X_{\rm \ci}\approx 10^{-3.9 \pm 0.1}$) we are able to get our flux density ratio-derived gas depletion timescales in line with IR-based estimates. 

\subsection{Extended \cii emission in DSFGs}

Now that we have determined that the \cii is extended in both galaxies, we compare our findings to similar studies in the literature. We first note that in the Appendix, we describe several tests of the robustness of the \cii measurement, by considering possible source plane factors such as the \cii-detected companion galaxy in \cite{cathey2024} and the AGN candidate in \cite{birkin2023}. Attempting to account for either of these additional components in the lens modeling process does not affect the main finding of extended \cii emission. 

The effective radius of \cii is about three times as extended as the dust continuum in SPT0418-47, and about twice as extended in SPT2147-50. This makes the \cii emission comparable in spatial extent with e.g. \cite{gullberg2018} who found that \cii was more extended than dust continuum by a factor of 2.1 $\pm$ 0.4 in four DSFGs from $z = 4.4-4.8$. It has been observed in the constituent galaxies of the SPT2349-56 protocluster system \citep{hill2022} that the mean half-light radius of \cii emission is 30 $\pm$ 20 \% larger than the half-light radius of FIR emission. The lensed system SPT0311-58 ($z = 6.9$) also features extended \cii emission compared to the dust continuum \citep{Spilker2022}.  \cite{amvrosiadis2024} similarly showed a difference in $r_{\rm \cii}$ vs. $r_{\rm 160}$ in SPT2147-50 using alternative modeling tools from this work, finding that the spatial extent of \cii ($r_{\rm e} = 2.67 \pm 0.07$ kpc) vs. dust continuum ($r_{\rm e} = 0.72 \pm 0.01$ kpc) differed by a factor of $\approx$ 3.7. Together, these works suggest that the extended \cii emission in SPT0418-47 and SPT2147-50 is not a surprising discovery in DSFGs. 

Beyond DSFGs, a literature review of high-redshift galaxies in \cite{gullberg2018} found a mean size ratio of \cii / 160$\mu$m dust continuum $=$ 1.6 $\pm$ 0.4. In the ALPINE-ALMA survey, \cite{fujimoto2020} found systematically extended \cii emission vs. rest-frame UV (via \textit{HST} imaging), with $r_{\rm e, \cii} / r_{\rm e, UV} \sim 2-3$. A growing body of observational work has explored the ubiquity of extended \cii vs. dust in high redshift galaxies, first by stacking \citep[e.g.][in ALPINE]{fujimoto2019, ginolfi2020}, which uncovered 10 kpc \cii ``haloes'', and eventually by direct observations \citep{fujimoto2020, akins2022, lambert2023, posses2024}. In luminous local galaxies, the \cii emission is more extended than the dust continuum \citep{Diaz2014}, but not in dwarf galaxies \citep{romano2024}, who caution that the lack of similar extent in \cii vs. dust continuum can be due to sensitivity effects and effects of the cosmic microwave background (CMB). 

There are a number of explanations for extended \cii in galaxies, the simplest being that \cii emission arises from the molecular, atomic, and ionized phase of gas in galaxies \citep[e.g.][]{pineda2013, croxall2017, vizgan22a, casavecchia2025}. Thus it is natural to expect that \cii, as a tracer of the cold gas, should be more extended \citep{tumlinson2017} than warm dust, which is closely tied to regions of newly formed stars that are heating it. A further possible explanation of extended \cii emission is outflowing gas, a hypothesis which agrees with theoretical works \citep{Pizzati2020, pizzati2023}. Recent work from the CRISTAL survey \citep{herreracamus2025} has found that extended \cii emission can be explained by merger activity \citep{posses2024}, but that extended \cii emission may not be as ubiquitous or as significant as expected in high-$z$ galaxies \citep{ikeda2025}. 

What we show in this work is that in our two DSFGs, the \cii is not only extended with respect to two bands of dust (160$\mu$m, 380$\mu$m continuum) -- which is not so unexpected -- but, importantly, that \cii is more extended than other tracers of star-forming gas. We thus caution that \cii, at least on resolved scales, may not be an ideal tracer of ongoing star-formation due to its vast spatial extent and evident ubiquity across all phases of the ISM. It is nevertheless utilized as a tracer of physical properties in high-redshift galaxies such as SFRs \citep[e.g.][]{lagache2018} and molecular gas masses \citep[e.g.][]{zanella2018}. 

\section{Conclusions} \label{sec:conc}

SPT0418-47 and SPT2147-50 are two lensed DSFGs at $z \approx 4$, which have been extensively studied over the preceding decade. In this work, we have performed a comprehensive investigation of both sources at $\approx$ 0.5\arc resolution via new and archival ALMA observations targeting dust continuum, \coseven, \citwo, \cii, and, in SPT0418-47, $p-$H$_2$O emission. In studying these lines, we examine some of the most commonly used star-formation and gas-mass tracers in the literature that are applied to high-redshift galaxies. These lines are all robustly detected at high signal-to-noise, with similar line widths and profiles in each individual galaxy.

In the image plane, we take advantage of the gravitationally lensed nature of both sources to study the spatially resolved Kennicutt-Schmidt relation. Utilizing conversion factors and log-linear scaling relations from the literature, we convert spatially resolved maps of line emission into maps of SFR and molecular gas surface densities. After fitting lines of best-fit to these points, we conclude that, for the most part, that many different combinations of observed emission lines are able to retrieve the Kennicutt-Schmidt relation, with six of the seven combinations retrieving a slope $m$ ranging from $1.00 \pm 0.03$ to $1.09 \pm 0.03$, all of which are in excellent agreement with recent work by \cite{kennicutt2021} on local starbursts ($m = 0.98 \pm 0.07$). After averaging the pixels to one point, we find that six of the seven combinations return gas depletion timescale estimates $\lesssim$ 200 Myr, also as expected for starburst galaxies. The odd one out in both cases involves \cii emission; \cii as a tracer of molecular gas results in $m = 1.20 \pm 0.03$ vs \coseven as a star formation tracer, and \cii as a SFR tracer vs. \citwo as a cold gas tracer yields gas depletion timescales $> 200$ Myr. This is interesting both due to the ubiquitous use of \cii luminosity as a tracer of both SFR and $M_{\rm mol}$, and because \coseven is expected to be a natural tracer of star-formation, just as \citwo is for molecular gas.

In the source plane, we utilize lens modeling tools to resolve the physical structure of both DSFGs. We take advantage of our multi-band observations to place excellent constraints on the foreground lens models. After assuming a shared source plane geometry across every band, we find that in both sources, the dust continuum is the most compact, followed by \coseven and \citwo, which arise from the same physical region in both galaxies. The water line observed in SPT0418-47 is as or more compact than the dust emission and shows evidence of being highly truncated spatially. In both galaxies, the \cii emission is significantly extended spatially, well beyond all other lines. We show in the Appendix that the companion galaxy in SPT0418-47 \citep{cathey2024}, and the AGN candidate in SPT2147-50 \citep{birkin2023}, cannot account for these extended profiles, and even with a different source plane profile for \cii emission we still recover spatially extended components. This is not a surprising finding when considering that other studies of SPT-selected DSFGs \citep{gullberg2018, hill2020, Spilker2022} have recovered extended \cii emission compared to dust continuum and/or FIR. 

When we consider our findings in both the image plane and source plane, we determine that the \citwo to \coseven flux density ratio is an attractive proxy for the gas depletion timescale $\tau_{\rm dep}$ in DSFGs. In tandem, these lines yield excellent agreement with the spatially resolved Kennicutt-Schmidt relation, and are physically coincident with one another compared to dust and \cii; they can be efficiently observed in a single tuning band with ALMA over a wide range of source redshifts, and the ratio of these lines is largely insensitive to magnification due to gravitational lensing. After combining equations from \cite{weiss2003} and \cite{lu2015}, we derive an expression for the gas depletion timescale that is dependent only on $X_{\rm \ci}$, $R_{\rm \ci}$, and the flux density ratio of \citwo to \coseven. This expression is most sensitive to $X_{\rm \ci}$, but will enable astronomers to obtain reasonable estimates of $\tau_{\rm dep}$  right off of an observed DSFG spectrum in the sub-millimeter window (via e.g. ALMA or APEX). 

We show in Figures \ref{fig:tdepvsz} and \ref{fig:comparison_tdep} that our expression retrieves values for $\tau_{\rm dep}$ that are as accurate as those derived from SED-fitting \citep{reuter2020}. The SED-derived $\tau_{\rm dep}$ requires many observations across multiple frequencies, compared to the flux density-derived $\tau_{\rm dep}$, which can be obtained easily with an hour or less of on-source time with ALMA. In order to get our gas depletion time estimates to be best-aligned with SED fitting, we require an implied $X_{\rm \ci}$ that is at the largest end of what has been published in the literature \citep{valentino2018}. 

An important caveat to note about this work is that the \ci/CO flux density ratio, as a proxy for $\tau_{\rm dep}$, that is developed in this paper has not been fully realized towards spatially resolved scales, and will require further work. This paper also highlights the unique strengths of ALMA in studying gravitationally lensed, high-redshift galaxies. The wealth of observable lines obtainable in these SPT-selected DSFGs, combined with the high spatial and spectral resolutions achievable with ALMA, provide unique insights into this population of galaxies and the physical properties of the high-redshift interstellar medium.

\section*{Acknowledgements} \label{sec:ack}
We thank the anonymous referee for their feedback, which improved this work. DV thanks Zachary D. Stone for insightful discussions on code which greatly improved the analysis, and extends gratitude to Marcel Neeleman, along with the entire staff of the North American ALMA Regional Center (NAARC) at the National Radio Astronomy Observatory (NRAO), for facilitating the start of this project via in-person support with data reduction. The National Radio Astronomy Observatory is a facility of the National Science Foundation operated under cooperative agreement by Associated Universities, Inc. The SPT is supported by the NSF through grant OPP-1852617. This work was partially supported by the Center for AstroPhysical Surveys (CAPS) at the National Center for Supercomputing Applications (NCSA), University of Illinois Urbana-Champaign. This material is based upon work supported by the National Science Foundation Graduate Research Fellowship under Grant No. DGE 21\~46756. MA is supported by FONDECYT grant number 1252054, and gratefully acknowledges support from ANID Basal Project FB210003 and ANID MILENIO NCN2024\_112. 

This paper makes use of the following ALMA data: ADS/JAO.ALMA \# 2016.1.01374.S, ADS/JAO.ALMA \# 2016.1.01499.S, ADS/JAO.ALMA \# 2018.1.01060.S, ADS/JAO.ALMA \# 2019.1.00471.S, ADS/JAO.ALMA \# 2021.1.00252.S. ALMA is a partnership of ESO (representing its member states), NSF (USA) and NINS (Japan), together with NRC (Canada) and NSC and ASIAA (Taiwan), in cooperation with the Republic of Chile. The Joint ALMA Observatory is operated by ESO, AUI/NRAO and NAOJ. This research has made use of NASA’s Astrophysics Data System and Astrophysics Data System Bibliographic Services. 

\vspace{5mm}
\facilities{ALMA}
\software{Python, CASA \cite{mcmullin2007}, CARTA \citep{comrie2021}, spectral-cube \citep{ginsburg2019} astropy \citep{astropy2022}, matplotlib, scipy, numpy, pandas, lenstronomy \citep{birrer2021}} 

\bibliography{sample631}{}
\bibliographystyle{aasjournal}
\FloatBarrier 
\begin{appendices}

\section{Best-fit lens models for SPT0418-47 and SPT2147-50}\label{sec:appa}

\begin{figure*}[h!]
    \centering
    \begin{subfigure}
        \centering
        \includegraphics[trim = 0 5cm 0 5cm, clip, width=\textwidth]{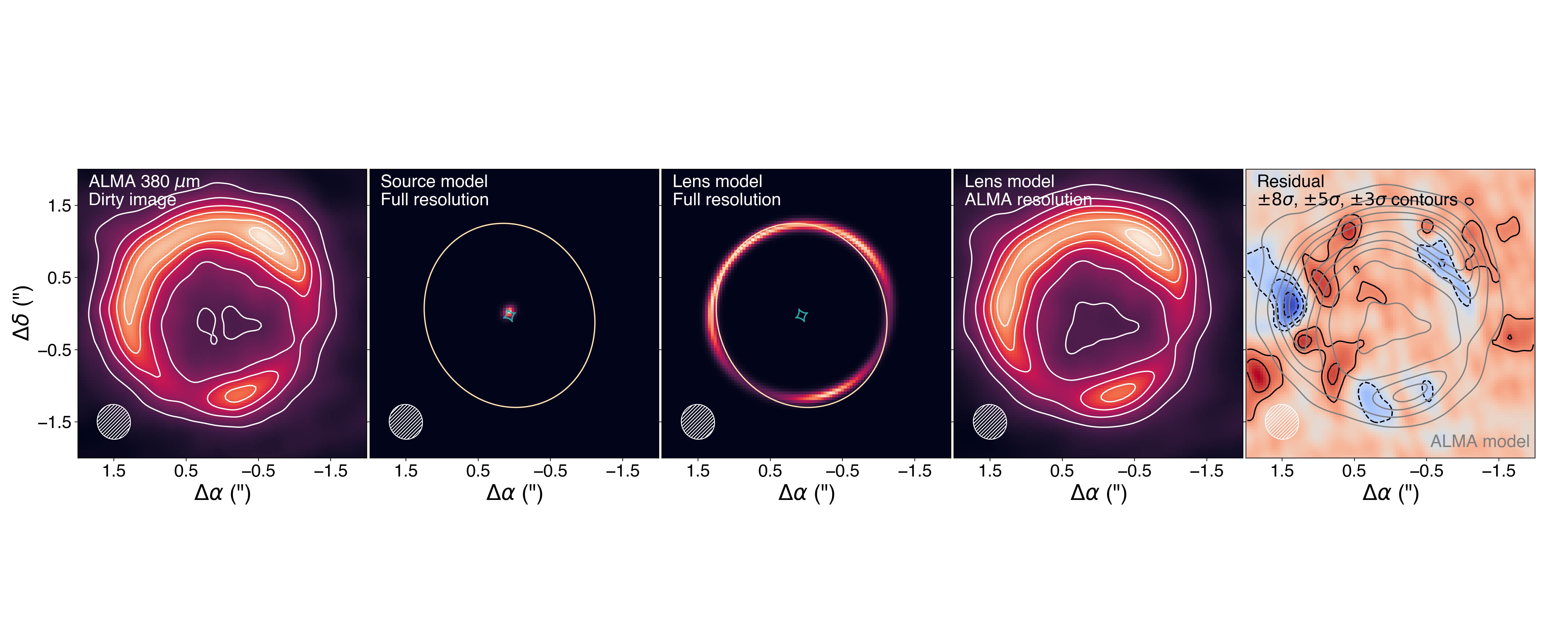}
    \end{subfigure}

    \begin{subfigure}
        \centering
        \includegraphics[trim = 0 5cm 0 5cm, clip, width=\textwidth]{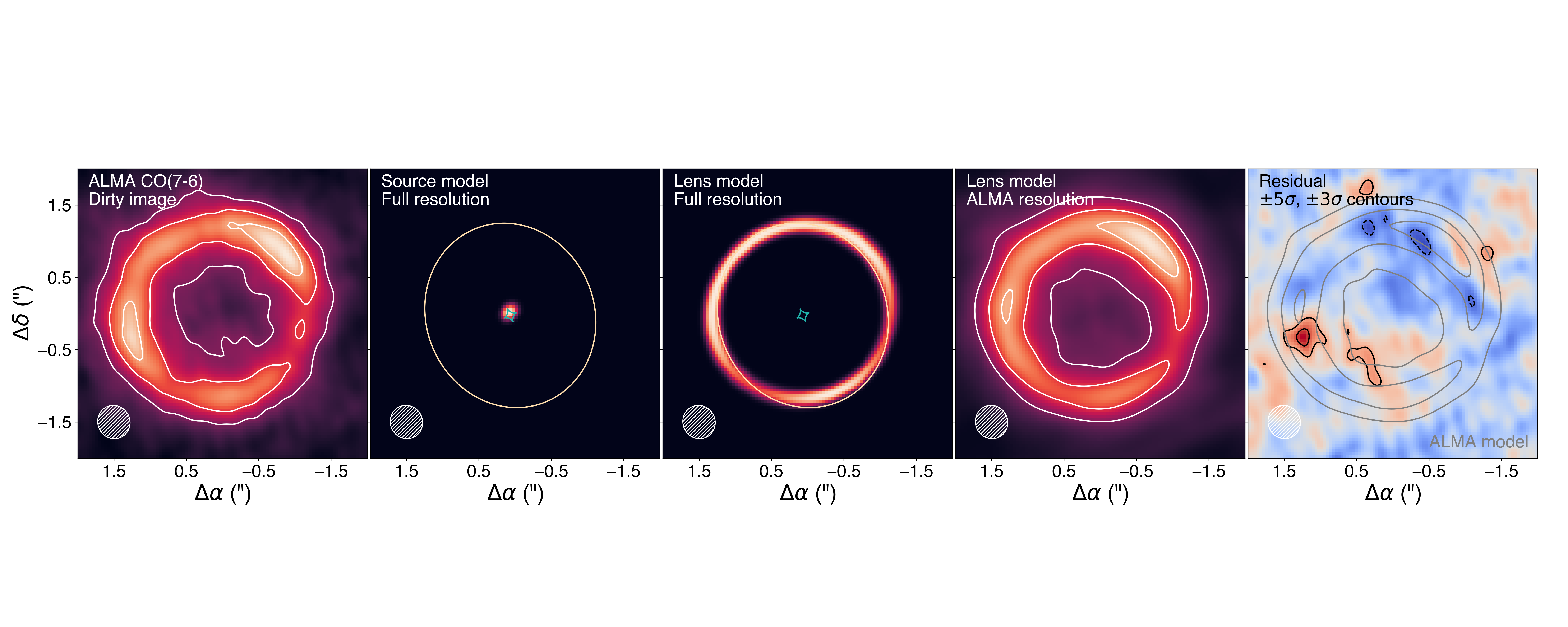}
    \end{subfigure}

    \begin{subfigure}
        \centering
        \includegraphics[trim = 0 5cm 0 5cm, clip, width=\textwidth]{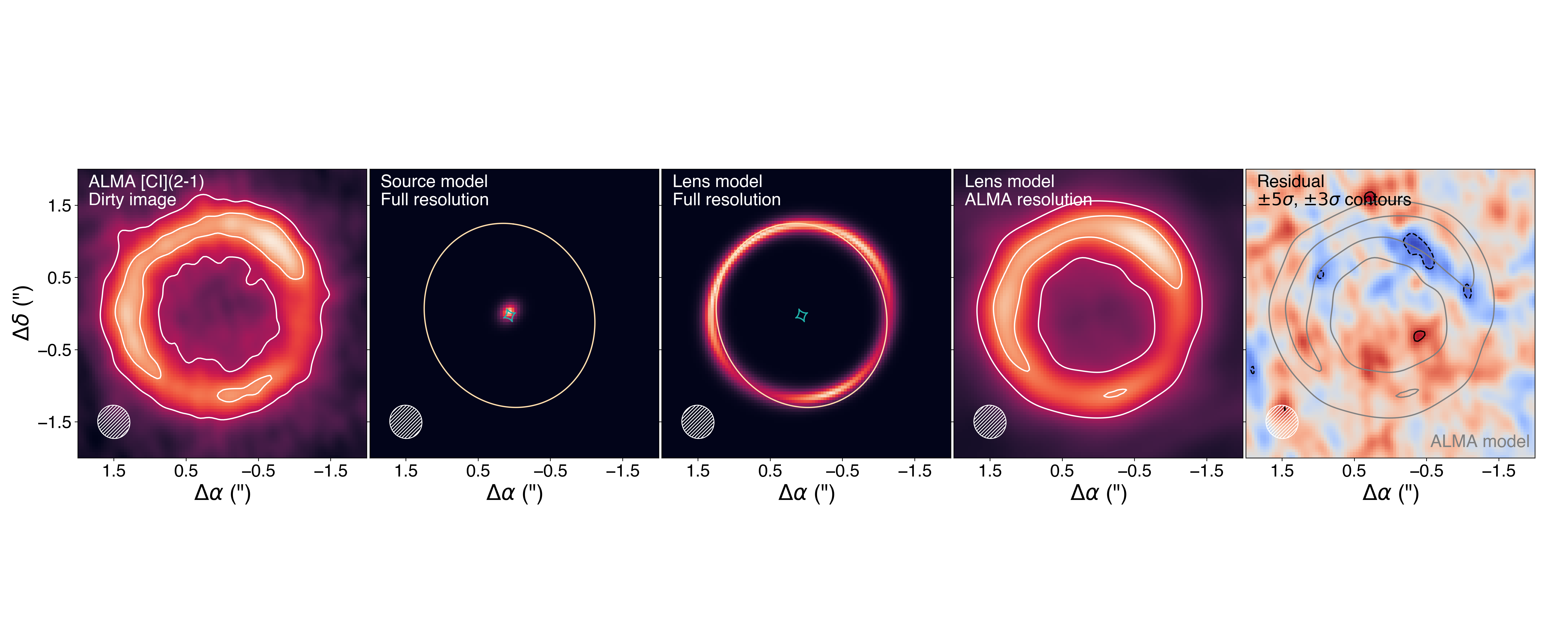}
    \end{subfigure}

    \caption{\textbf{Selection of lens models for SPT0418-47.} Each subfigure shows, from left to right: \textbf{a)} The dirty (un-CLEANed) image of emission; \textbf{b)} the best-fit source plane model, with overlaid caustic (cyan) and critical curve (light gold); \textbf{c)} the lens model, i.e. ray-tracing of the source-plane model across the foreground mass profile, at the native pixel resolution; \textbf{d)} The same lens model, now convolved to the ALMA data resolution, \textbf{e)} the residual, i.e. dirty image minus best-fit lens model. The beam is visualized as a hatched ellipse in the bottom left corner; all dirty images were beam-matched to the \coseven restoring beam to enable direct comparison. From top to bottom, we show 380 $\mu$m continuum ($\sigma = 5.2$ $\mu$Jy/beam), \coseven ($\sigma = 24$ $\mu$Jy/beam km s$^{-1}$), and \citwo emission ($\sigma = 25$ $\mu$Jy/beam km s$^{-1}$). Contours (white lines) are generated as integer multiples of 25$\sigma$, 10$\sigma$, and 10$\sigma$ from top to bottom, respectively. The native pixel resolution of each subpanel is 0.04\arc.}
    \label{fig:SPT0418-47_lens_models_pt1}
\end{figure*}

\begin{figure*}[h!]
    \centering

    \begin{subfigure}
        \centering
        \includegraphics[trim = 0 5cm 0 5cm, clip, width=\textwidth]{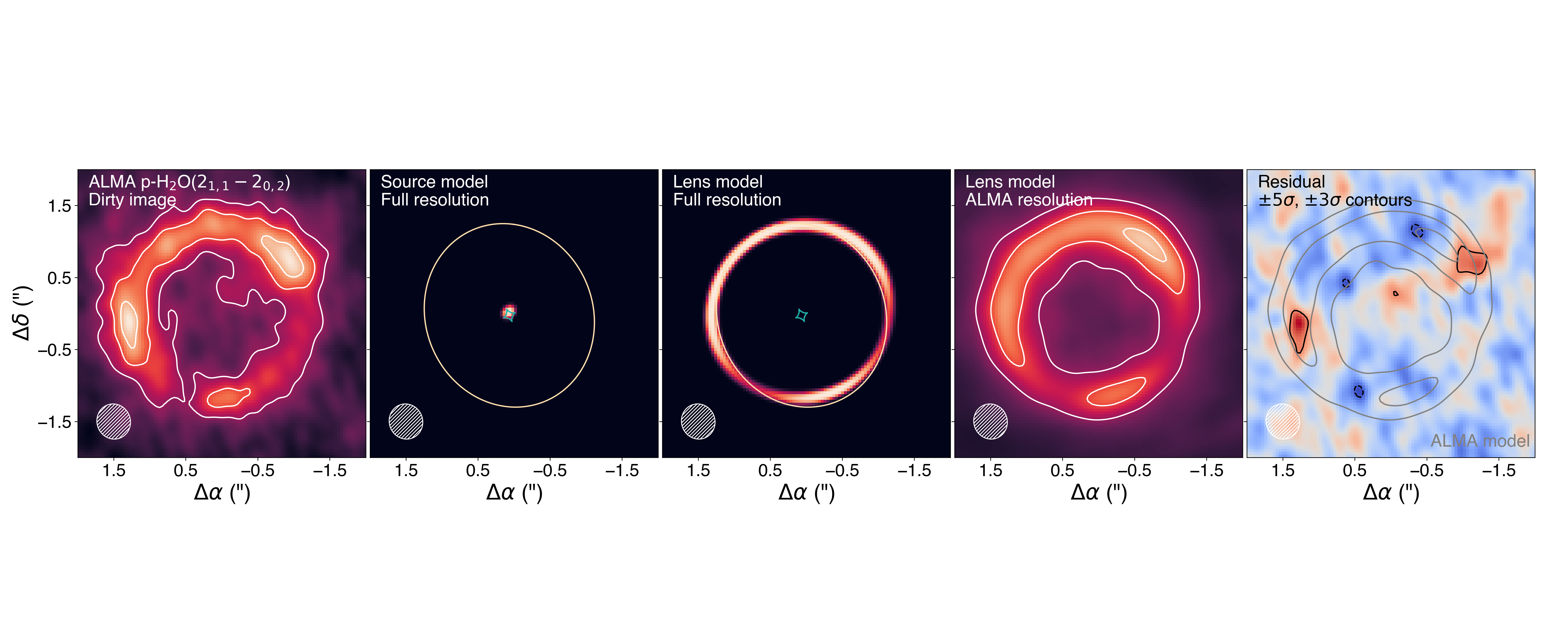}
    \end{subfigure}
    
    \begin{subfigure}
        \centering
        \includegraphics[trim = 0 5cm 0 5cm, clip, width=\textwidth]{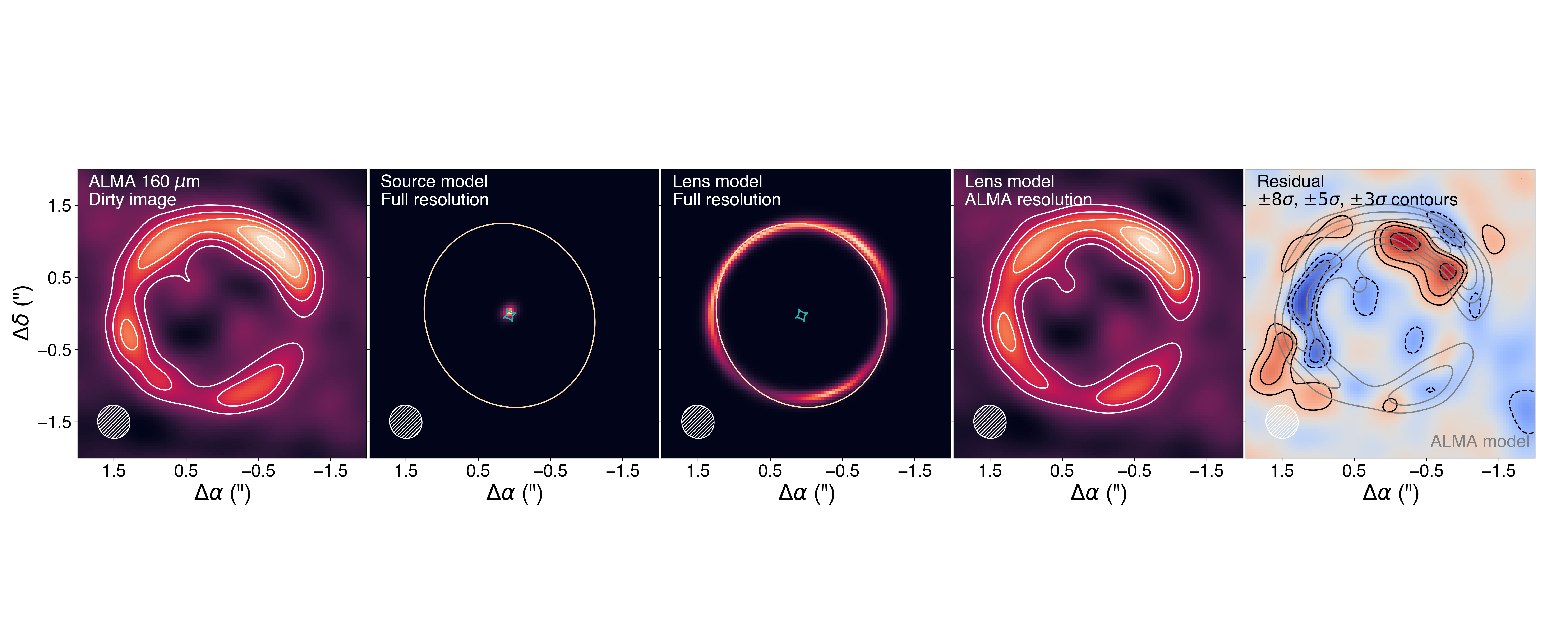}
    \end{subfigure}

    \begin{subfigure}
        \centering
        \includegraphics[trim = 0 5cm 0 5cm, clip, width=\textwidth]{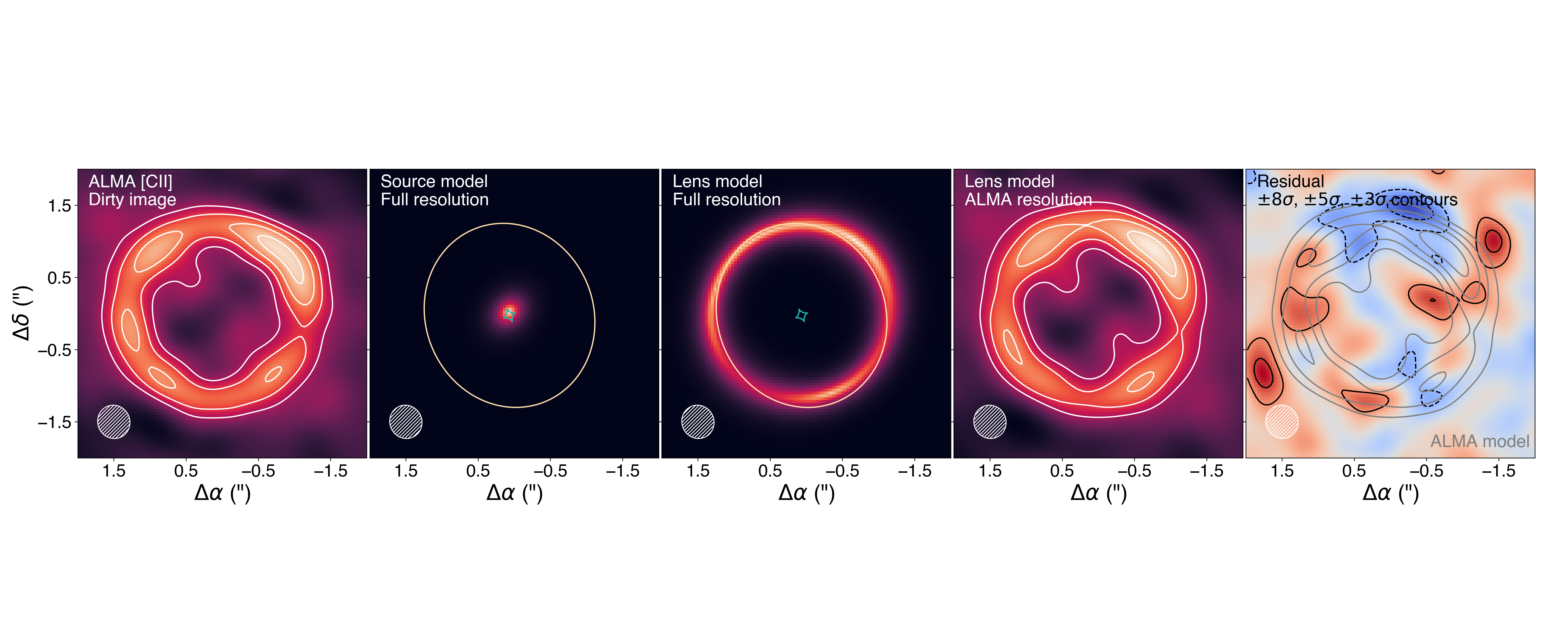}
    \end{subfigure}
    
    \caption{\textbf{Selection of lens models for SPT0418-47, continued.} From top to bottom, we show $p-$H$_2$O($2_{1,1}-2_{0,2}$) emission ($\sigma = 19$ $\mu$Jy/beam km s$^{-1}$), 160 $\mu$m dust continuum ($\sigma = 80$ $\mu$Jy/beam), and \cii emission ($\sigma = 230$ $\mu$Jy/beam km s$^{-1}$). Contours (white lines) are generated as integer multiples of 5$\sigma$, 20$\sigma$, and 10$\sigma$ from top to bottom, respectively. The native pixel resolution of each subpanel is 0.04\arc.}
    \label{fig:SPT0418-47_lens_models_pt2}
\end{figure*}

\begin{figure*}[h!]
    \centering
    \begin{subfigure}
        \centering
        \includegraphics[trim = 0 5cm 0 5cm, clip, width=\textwidth]{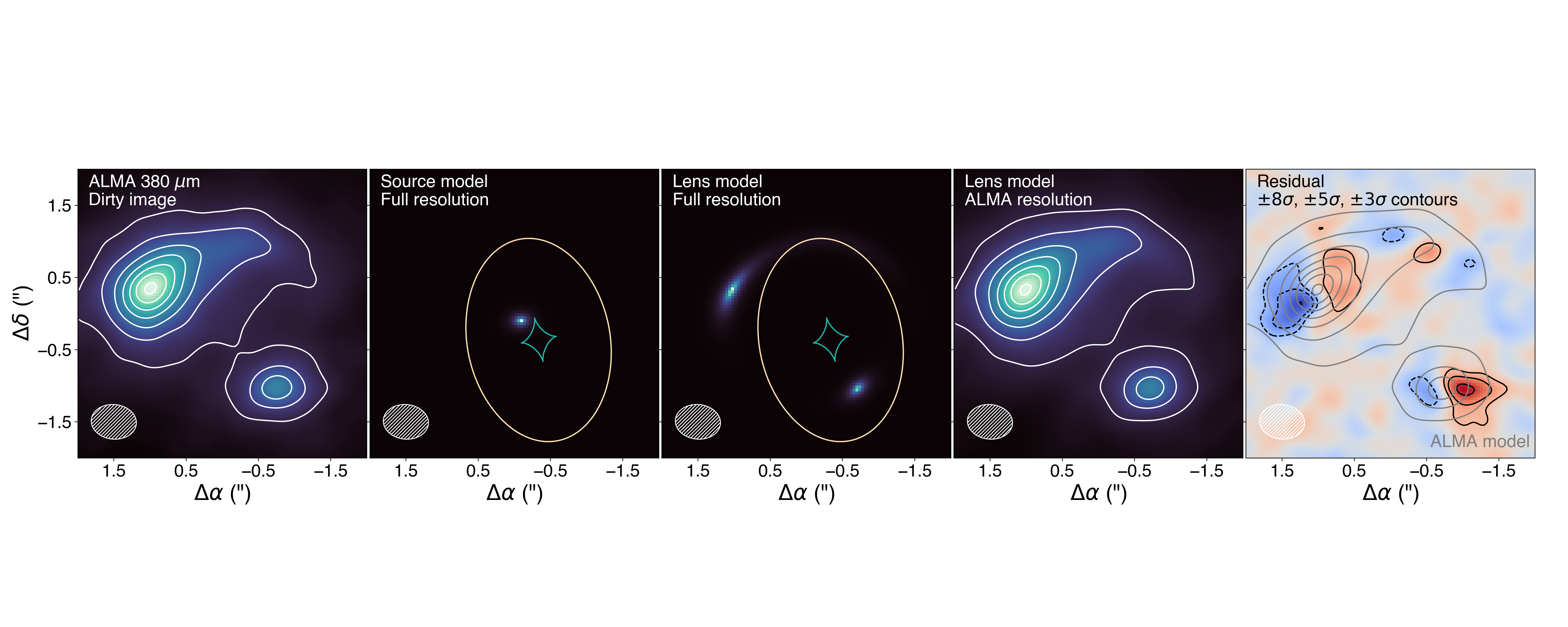}
    \end{subfigure}

    \begin{subfigure}
        \centering
        \includegraphics[trim = 0 5cm 0 5cm, clip, width=\textwidth]{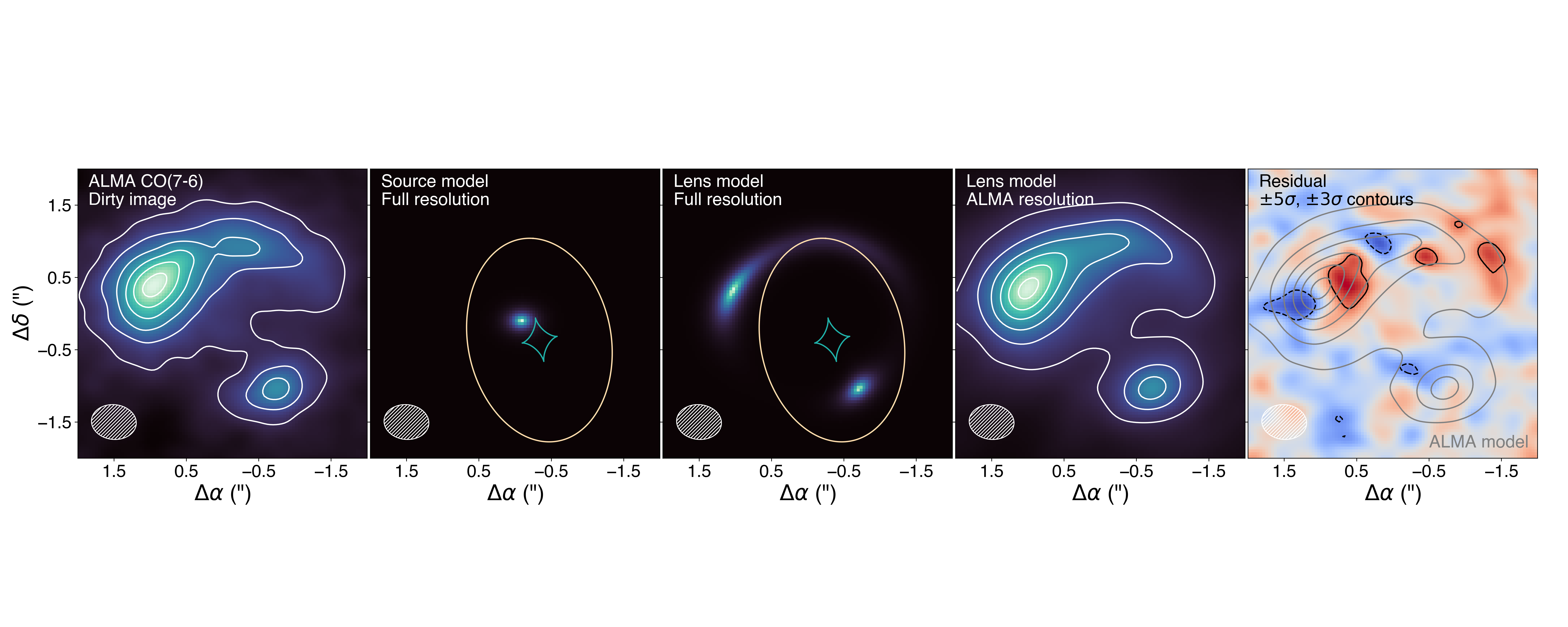}
    \end{subfigure}

    \begin{subfigure}
        \centering
        \includegraphics[trim = 0 5cm 0 5cm, clip, width=\textwidth]{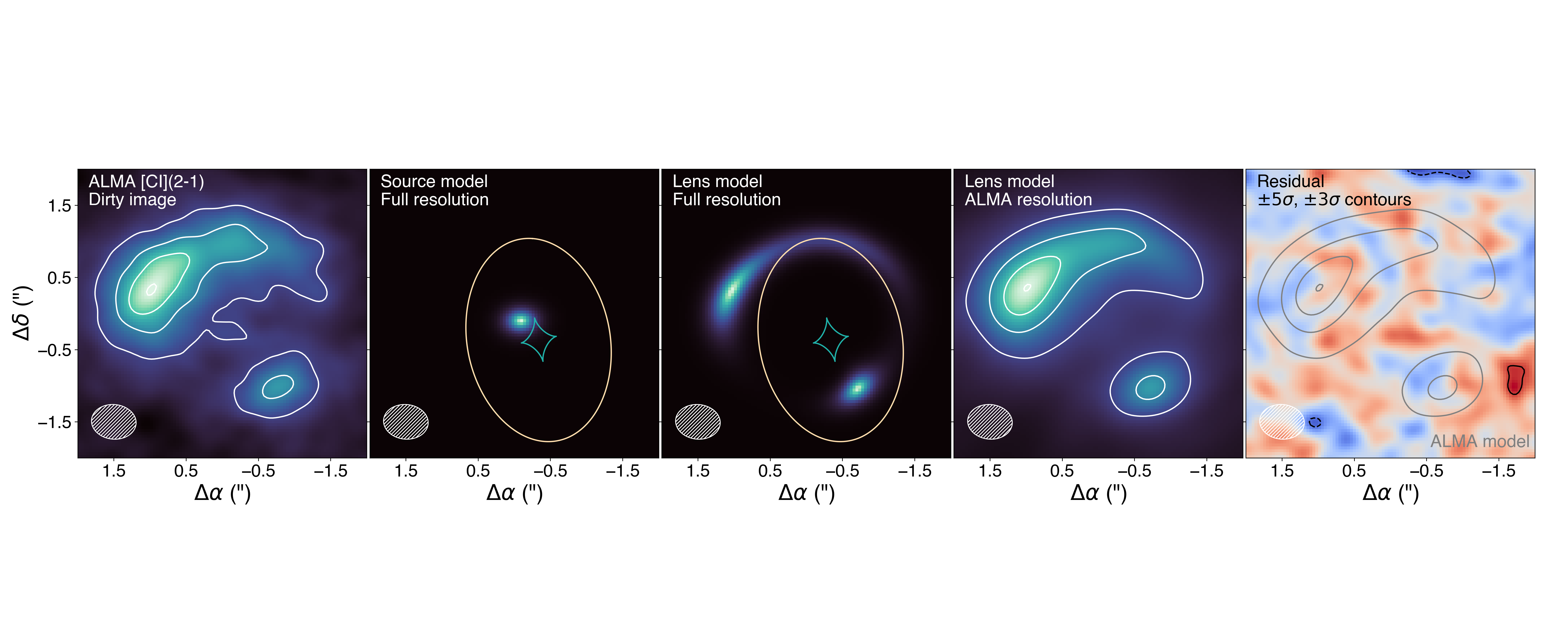}
    \end{subfigure}

    \caption{\textbf{Selection of lens models for SPT2147-50.} Same as in Figures \ref{fig:SPT0418-47_lens_models_pt1} and \ref{fig:SPT0418-47_lens_models_pt2}. From top to bottom, we show 380 $\mu$m continuum ($\sigma = 11$ $\mu$Jy/beam), \coseven ($\sigma = 28$ $\mu$Jy/beam km s$^{-1}$), and \citwo emission ($\sigma = 28$ $\mu$Jy/beam km s$^{-1}$). Contours (white lines) are generated as integer multiples of 25$\sigma$, 10$\sigma$, and 10$\sigma$ from top to bottom, respectively. The native pixel resolution of each subpanel is 0.04\arc.}
    \label{fig:SPT2147-50_lens_models_pt1}
\end{figure*}

\begin{figure*}[h!]
    \centering
    
    \begin{subfigure}
        \centering
        \includegraphics[trim = 0 5cm 0 5cm, clip, width=\textwidth]{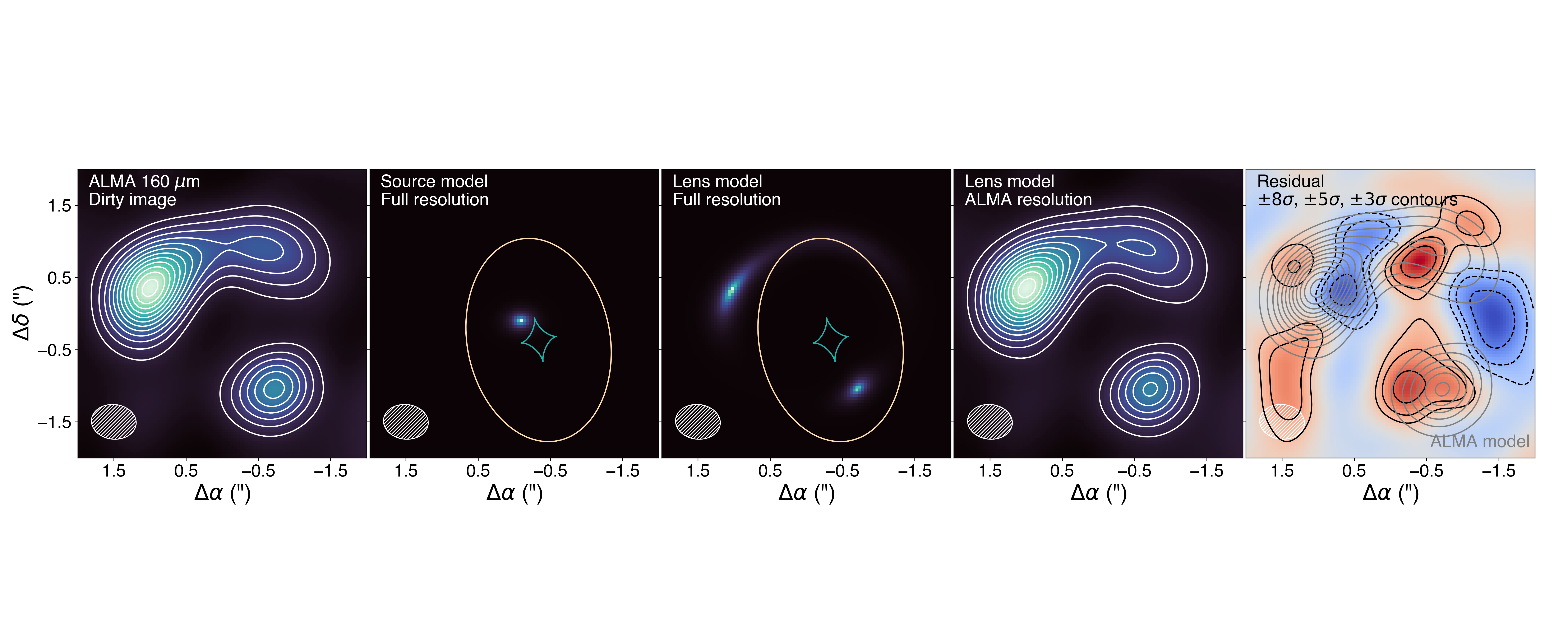}
    \end{subfigure}

    \begin{subfigure}
        \centering
        \includegraphics[trim = 0 5cm 0 5cm, clip, width=\textwidth]{SPT2147-50_CII_DMR_plot.png}
    \end{subfigure}

    \caption{\textbf{Selection of lens models for SPT2147-50, continued.} We show 160 $\mu$m continuum ($\sigma = 70$ $\mu$Jy/beam) on top, and \cii emission ($\sigma = 57$ $\mu$Jy/beam km s$^{-1}$) on bottom. Contours (white lines) are generated as integer multiples of 30$\sigma$, and 20$\sigma$ for the top and bottom subfigures, respectively. The native pixel resolution of each subpanel is 0.04\arc.}
    \label{fig:SPT2147-50_lens_models_pt2}
\end{figure*}

\begin{table}[h!]
    \centering
    \begin{tabular}{lccc}
        \toprule
        \textbf{Parameters} & \textbf{Median} & \textbf{Best-fit} & \textbf{Priors} \\
        \midrule
        $\theta_E$ [\arc] & 1.2247$^{+0.0005}_{-0.0005}$ & 1.2246 & (0.5, 2.0) \\
        $e_{1, \ \rm{lens}}$ & -0.038$^{+0.003}_{-0.003}$ & -0.037 & (-1, 1) \\
        $e_{2, \ \rm{lens}}$ & -0.030$^{+0.003}_{-0.003}$ & -0.032 & (-1, 1) \\
        $\Delta x_{\rm{lens}}$ [\arc] & -0.0654$^{+0.0009}_{-0.0009}$ & -0.0650 & (-0.5, 0.5) \\
        $\Delta y_{\rm{lens}}$ [\arc] & -0.0258$^{+0.0014}_{-0.0014}$ & -0.0261 & (-0.5, 0.5) \\
        $\gamma_{1}$ & 0.0004$^{+0.0014}_{-0.0014}$ & 0.0005 & (-0.5, 0.5) \\
        $\gamma_{2}$ & 0.0068$^{+0.0015}_{-0.0014}$ & 0.0058 & (-0.5, 0.5) \\
        $r_{e, \ 380\ \mu m}$ [\arc] & 0.0625$^{+0.0013}_{-0.0013}$ & 0.0626 & (0.005, 0.5) \\
        $n_{\ 380\ \mu m}$ & 0.63$^{+0.03}_{-0.03}$ & 0.63 & (0.05, 2)\\
        $r_{e, \ \rm{\scriptsize \coseven}}$ [\arc] & 0.086$^{+0.006}_{-0.006}$ & 0.080 &  (0.005, 0.5) \\
        $n_{\rm{\scriptsize \coseven}}$ & 0.45$^{+0.09}_{-0.09}$ & 0.38 & (0.05, 3) \\
        $r_{e, \ \rm{\scriptsize \citwo}}$ [\arc] & 0.112$^{+0.010}_{-0.009}$ & 0.108 & (0.005, 0.5) \\
        $n_{\rm{\scriptsize \citwo}}$ & 0.77$^{+0.13}_{-0.12}$ & 0.73 & (0.05, 3) \\
        $r_{e, \ \rm{H_2O}}$ [\arc] & 0.063$^{+0.008}_{-0.008}$ & 0.061 & (0.005, 0.5) \\
        $n_{\rm{H_2O}}$ & 0.30$^{+0.15}_{-0.09}$ & 0.27 & (0.05, 3) \\
        $r_{e, \ 160\ \mu m}$ [\arc] & 0.079$^{+0.002}_{-0.002}$ & 0.079 & (0.005, 0.5)  \\
        $n_{160\ \mu m}$ & 0.73$^{+0.05}_{-0.05}$ & 0.72 & (0.05, 3) \\
        $r_{e, \ \rm{\cii}}$ [\arc] & 0.223$^{+0.010}_{-0.009}$ & 0.220 & (0.005, 0.5) \\
        $n_{\rm{\cii}}$ & 0.96$^{+0.07}_{-0.06}$ & 0.95 & (0.05, 3) \\
        $e_{1, \ \rm source}$ & -0.014$^{+0.006}_{-0.006}$ & -0.008 & (-1, 1) \\
        $e_{2, \ \rm source}$ & 0.107$^{+0.004}_{-0.004}$ & 0.110 & (-1, 1) \\
        $\Delta x_{\rm{source}}$ [\arc] & -0.0755$^{+0.0009}_{-0.0009}$ & -0.0751 & (-0.5, 0.5) \\
        $\Delta y_{\rm{source}}$ [\arc] & 0.0211$^{+0.0007}_{-0.0007}$ & 0.0210 & (-0.5, 0.5) \\
        \bottomrule
    \end{tabular}
    \caption{Lens modeling results, via MCMC, in SPT0418-47, fitting 380$\mu$m dust continuum, \coseven, \citwo, $p$-H$_2$O, 160 $\mu$m dust continuum, and \cii to a shared source geometry (i.e. same $e_1, e_2$, $\Delta x, \Delta y$ in the source plane). 
    }
    \label{tab:parameters_0418}
\end{table}

\begin{table}[h!]
    \centering
    \begin{tabular}{lccc}
        \toprule
        \textbf{Parameters} & \textbf{Median} & \textbf{Best-fit} & \textbf{Priors} \\
        \midrule
        $\theta_E$ [\arc] & 1.1851$^{+0.0011}_{-0.0011}$ & 1.1848 & (1.1, 1.25) \\
        $e_{1, \ \rm{lens}}$ & -0.181$^{+0.006}_{-0.006}$ & -0.178 & (-1, 1) \\
        $e_{2, \ \rm{lens}}$ & -0.067$^{+0.005}_{-0.005}$ & -0.067 & (-1, 1) \\
        $\Delta x_{\rm{lens}}$ [\arc] & 0.336$^{+0.002}_{-0.002}$ & 0.337 & (-0.5, 0.5) \\
        $\Delta y_{\rm{lens}}$ [\arc] & -0.367$^{+0.003}_{-0.003}$ & -0.367 & (-0.5, 0.5) \\
        $\gamma_{1}$ & -0.014$^{+0.004}_{-0.004}$ & -0.012 & (-0.5, 0.5) \\
        $\gamma_{2}$ & -0.002$^{+0.003}_{-0.003}$ & -0.002 & (-0.5, 0.5) \\
        $r_{e, \ 380\ \mu m}$ [\arc] & 0.1171$^{+0.0014}_{-0.0014}$ & 0.1163 & (0.01, 1) \\
        $n_{\ 380\ \mu m}$ & 1.26$^{+0.05}_{-0.05}$ & 1.28 & (0.05, 2)\\
        $r_{e, \ \rm{\scriptsize \coseven}}$ [\arc] & 0.154$^{+0.003}_{-0.003}$ & 0.152 &  (0.01, 1) \\
        $n_{\rm{\scriptsize \coseven}}$ & 0.95$^{+0.07}_{-0.07}$ & 0.98 & (0.05, 2) \\
        $r_{e, \ \rm{\scriptsize \citwo}}$ [\arc] & 0.174$^{+0.005}_{-0.005}$ & 0.173 & (0.01, 1) \\
        $n_{\rm{\scriptsize \citwo}}$ & 0.81$^{+0.09}_{-0.08}$ & 0.78 & (0.05, 2) \\
        $r_{e, \ 160\ \mu m}$ [\arc] & 0.1463$^{+0.0015}_{-0.0014}$ & 0.1462 & (0.01, 1)  \\
        $n_{160\ \mu m}$ & 1.26$^{+0.03}_{-0.03}$ & 1.28 & (0.05, 2) \\
        $r_{e, \ \rm{\cii}}$ [\arc] & 0.276$^{+0.004}_{-0.003}$ & 0.274 & (0.01, 1) \\
        $n_{\rm{\cii}}$ & 0.49$^{+0.02}_{-0.02}$ & 0.48 & (0.05, 2) \\
        $e_{1, \ \rm source}$ & 0.139$^{+0.007}_{-0.007}$ & 0.135 & (-1, 1) \\
        $e_{2, \ \rm source}$ & 0.011$^{+0.008}_{-0.008}$ & 0.013 & (-1, 1) \\
        $\Delta x_{\rm{source}}$ [\arc] & 0.088$^{+0.002}_{-0.002}$ & 0.089 & (-0.5, 0.5) \\
        $\Delta y_{\rm{source}}$ [\arc] & -0.1010$^{+0.0012}_{-0.0012}$ & -0.1011 & (-0.5, 0.5) \\
        \bottomrule
    \end{tabular}
    \caption{Same as Table \ref{tab:parameters_0418}, but for SPT2147-50. 
    }
    \label{tab:parameters_2147}
\end{table}

\begin{table}[h!]
    \centering
    \begin{tabular}{lcc}
        \toprule
        \textbf{Magnifications} & \textbf{SPT0418-47} & \textbf{SPT2147-50} \\
        \midrule
        $\mu_{\rm 380\,\mu m}$ & 41.7$^{+1.0}_{-0.8}$ & 6.38$^{+0.07}_{-0.08}$ \\
        $\mu_{\rm \scriptsize \coseven}$ & 34.4$^{+1.8}_{-1.6}$ & 6.82$^{+0.09}_{-0.08}$ \\
        $\mu_{\rm [C\,\textsc{i}](2-1)}$ & 28.6$^{+1.7}_{-1.6}$ & 7.04$^{+0.10}_{-0.10}$ \\
        $\mu_{\rm H_2 O}$ & 41.9$^{+1.8}_{-2.6}$ & \nodata \\
        $\mu_{\rm 160\,\mu m}$ & 36.4$^{+0.8}_{-0.8}$ & 6.62$^{+0.07}_{-0.07}$ \\
        $\mu_{\rm [C\,\textsc{ii}]}$ & 16.6$^{+0.5}_{-0.5}$ & 7.32$^{+0.09}_{-0.10}$ \\
        \bottomrule
    \end{tabular}
    \caption{Magnification factors ($\mu$) due to gravitational lensing for each modeled band.}
    \label{tab:magnifications}
\end{table}

\section{Searching for the \cii-detected companion in SPT0418-47}

It has been shown in the recent literature that SPT0418-47 has a companion galaxy which is detected both in \textit{JWST} imaging and in \cii \citep{peng2023, cathey2024}. This companion, named SPT0418-47B, lies roughly at a distance $\Delta \alpha, \Delta \delta == 0.6\arc, -0.3\arc$ away from the center of the source plane emission. It could reasonably be argued that the extended \cii shown in Figure \ref{fig:SPT0418-47_profiles} is a consequence of a \cii-bright companion alongside the main galaxy, which would require more spatially extended emission if only a single S\`ersic component is fitted in the source plane. In this appendix, we investigate what happens when we add a second component.

\begin{figure*}
    \centering
    \includegraphics[trim = 0 5cm 0 5cm, clip, width=\textwidth]{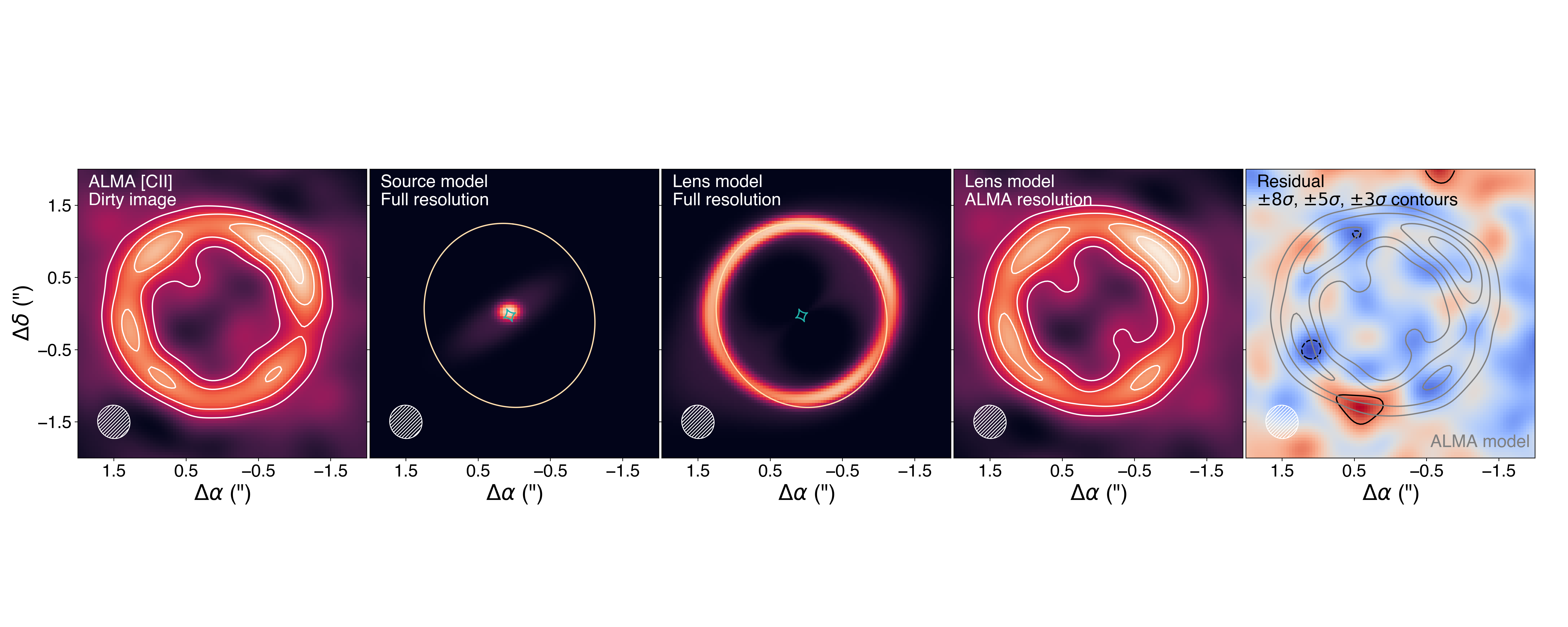}
    \caption{Two-component fit to \cii emission in SPT0418-47.}
    \label{fig:SPT0418-47_lens_models_companion_CII}
\end{figure*}

We first consider a holistic fit of two S\`ersic components to the data, ignoring the known location of the companion. To simplify our investigation, we apply the best-fit mass model, as constrained by six bands of ALMA data, so that our only parameters of interest are of those describing a 2D S\`ersic profile. We also do not constrain these profiles to share profile ellipticities with the other observables available (e.g. \coseven, \citwo, etc.). The resulting best-fit model recovers a central \cii profile which has a similar spatial extent to the \coseven and \citwo emission ($R_e = 0.11 \pm 0.01\arc$). The secondary profile, notably, is significantly more extended with $R_e = 0.38 \pm 0.04\arc$. The ellipcity is so extreme that the edges of the profile, as shown in Figure \ref{fig:SPT0418-47_lens_models_companion_CII}, are nearly touching the lensing critical curve along its semi-minor axis. This fit also reduces the maximum residuals from $>\pm5 \sigma$ to $\approx \pm3\sigma$, which indicates a better description of the data. This holistic fit does not recover a central source of emission at the location of the companion, and indicates that, even with a compact source of \cii tracing the same regions as \citwo and \coseven, a large, extended region of \cii is required to best characterize the data. Following this test, we attempted to confine the second component to lie only within the bottom-left quadrant (where the companion actually appears, both in the source and image planes). However, despite several attempts, the chains did not converge upon well-parameterized posterior distributions, nor did they retrieve a realistic solution during lens modeling.

\begin{table}[]
    \centering
    \begin{tabular}{lccc}
        \toprule
        \textbf{Parameters} & \textbf{Component 1} & \textbf{Component 2}  \\
        \midrule
        $R_{e, \ \rm{\cii}}$ [\arc] & 0.11$^{+0.01}_{-0.01}$ & 0.38$^{+0.04}_{-0.04}$ \\
        $n_{\rm{\cii}}$ & 0.43$^{+0.09}_{-0.10}$ & 0.40$^{+0.12}_{-0.11}$   \\
        $e_{1, \ \rm source}$ & -0.12$^{+0.02}_{-0.02}$ & 0.20$^{+0.03}_{-0.03}$  \\
        $e_{2, \ \rm source}$ & 0.01$^{+0.02}_{-0.02}$ & 0.49$^{+0.05}_{-0.05}$  \\
        $\Delta x_{\rm{source}}$ [\arc] & -0.073$^{+0.004}_{-0.004}$ & -0.17$^{+0.04}_{-0.05}$  \\
        $\Delta y_{\rm{source}}$ [\arc] & 0.024$^{+0.004}_{-0.003}$ & -0.06$^{+0.02}_{-0.03}$  \\
        \bottomrule
    \end{tabular}
    \caption{Two-component fit to SPT0418-47 \cii emission, after fixing the best-fit mass model from Table \ref{tab:parameters_0418}.
    }
    \label{tab:parameters_0418_companion}
\end{table}

\section{Further evidence for an active galactic nucleus in SPT2147-50?}

In SPT2147-50, we perform a similar multi-profile test in modeling the \cii emission to determine whether a bright core is affecting the extended \cii profile which emerges from the model. Again, after fixing our best-fit mass model of SPT2147-50 from the five-band fit (see Appendix \ref{sec:appa}) we first fit a double-component model, motivated by strong evidence for an AGN as presented in \cite{birkin2023}. The aim of this test is to see whether the \cii emission recovers a profile that is in agreement with a core and extended structure, so we fit a shared center position and ellipticity for both 2D profiles. The model parameter space, as explored by MCMC, does not converge upon a well-behaved solution; the S\`ersic index of the core component $n_{\rm core} = 0.06^{+0.03}_{-0.01}$ sits right along the lower bound of the prior ($n = 0.05$), showing very strong degeneracy with the effective radius $r_{\rm e, core} = 0.15^{+0.02}_{-0.03}$$\arc$. The extremely small S\`ersic index corresponding to a $\approx 1$ kpc region could indicate the presence of an AGN; unfortunately, the modification made to \texttt{lenstronomy} as described in \cite{zhang2025} does not permit for the modeling of a point source for interferometric-based images, which would help verify whether this degeneracy is evidence for an AGN. The extended second component in this fit was well-described by priors, with $r_{\rm e, ext} = 0.39^{+0.03}_{-0.02}$$\arc$ and $n_{\rm ext} = 0.99 \pm 0.10$. This corresponds to a spatial extent of $2.84^{+0.22}_{-0.15}$ kpc, in excellent agreement with what was found in \cite{amvrosiadis2024} by fitting a S\`ersic profile to their modeled \cii emission in SPT2147-50 ($r_e = 2.67 \pm 0.07$ kpc). 

\begin{figure*}
    \centering
    \includegraphics[width=\textwidth]{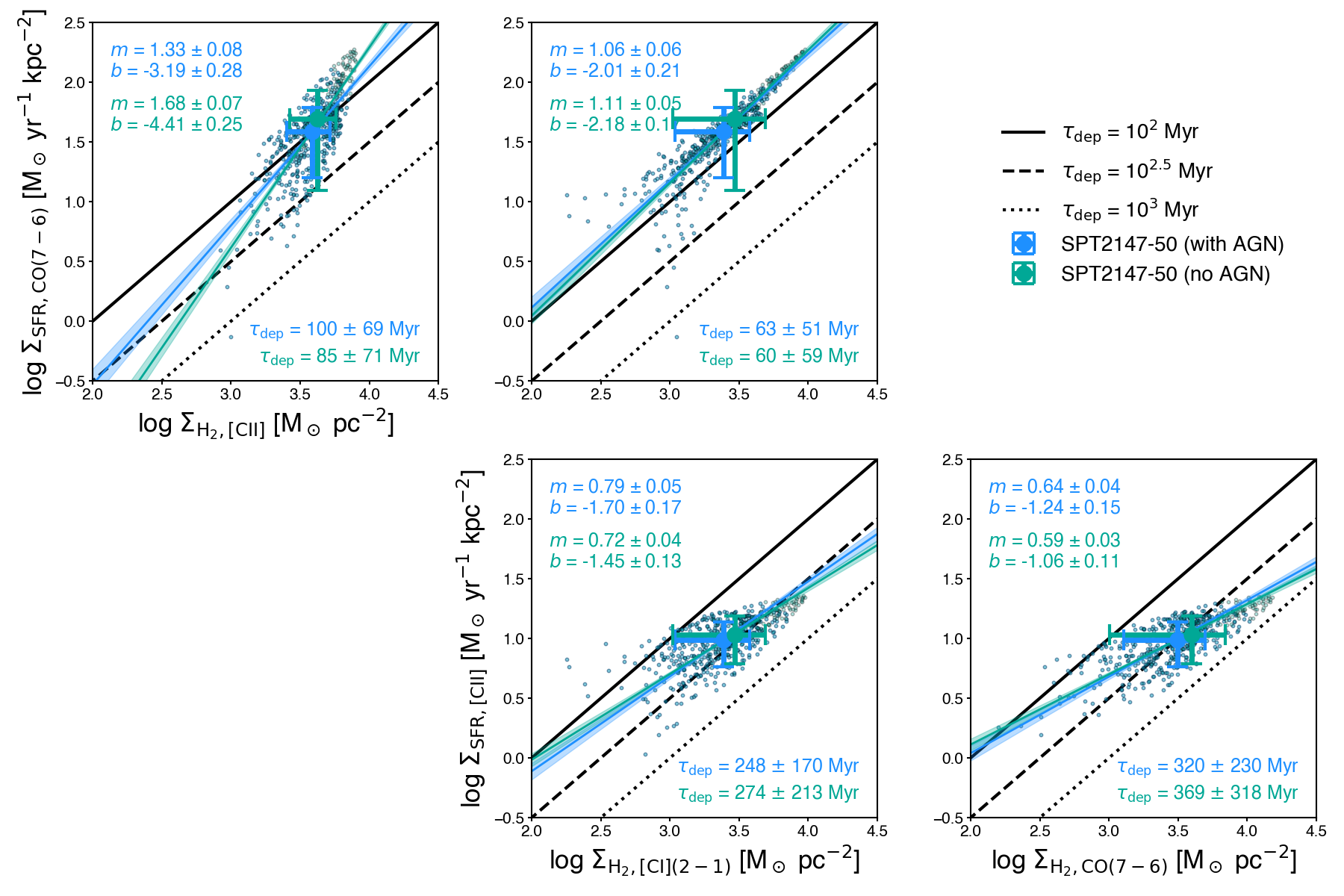}
    \caption{Spatially resolved Kennicutt-Schmidt relation, including (blue) and excluding (teal) the pixels which correspond to the speculated AGN in \cite{birkin2023}. In the right panels the effect of including or removing the AGN does not effect change in the fits beyond what is permitted in the uncertainties. The fit of \coseven vs. \cii as spatially resolved SFR vs. molecular gas surface density is strongly affected by removing the AGN.}
    \label{fig:SPT2147-50_KS_law_AGN}
\end{figure*}

We also tested how Figure \ref{fig:ks_law} might be affected by including or removing the pixels affected by the speculated AGN from \cite{birkin2023}, by redoing the fits for SPT2147-50 with and without the AGN-dominated pixels. This test is shown in Figure \ref{fig:SPT2147-50_KS_law_AGN}. When using \citwo as a gas mass tracer vs. \coseven and \cii as a SFR tracer, we do not see a clear difference in the slope of the fitted Kennicutt-Schmidt relation, regardless of the AGN. When studying \cii vs. \coseven as SFR vs. molecular gass mass, we find that removing the AGN does drive the best-fit slope higher, from $m = 1.33 \pm 0.08$ to $m = 1.68 \pm 0.07$. However, we have already shown in \ref{fig:ks_law} that this is, by far, the worst combination of observables to use for probing the Kennicutt-Schmidt relation in DSFGs like the ones studied in this work. 

\begin{figure*}
    \centering
    \includegraphics[width=\textwidth]{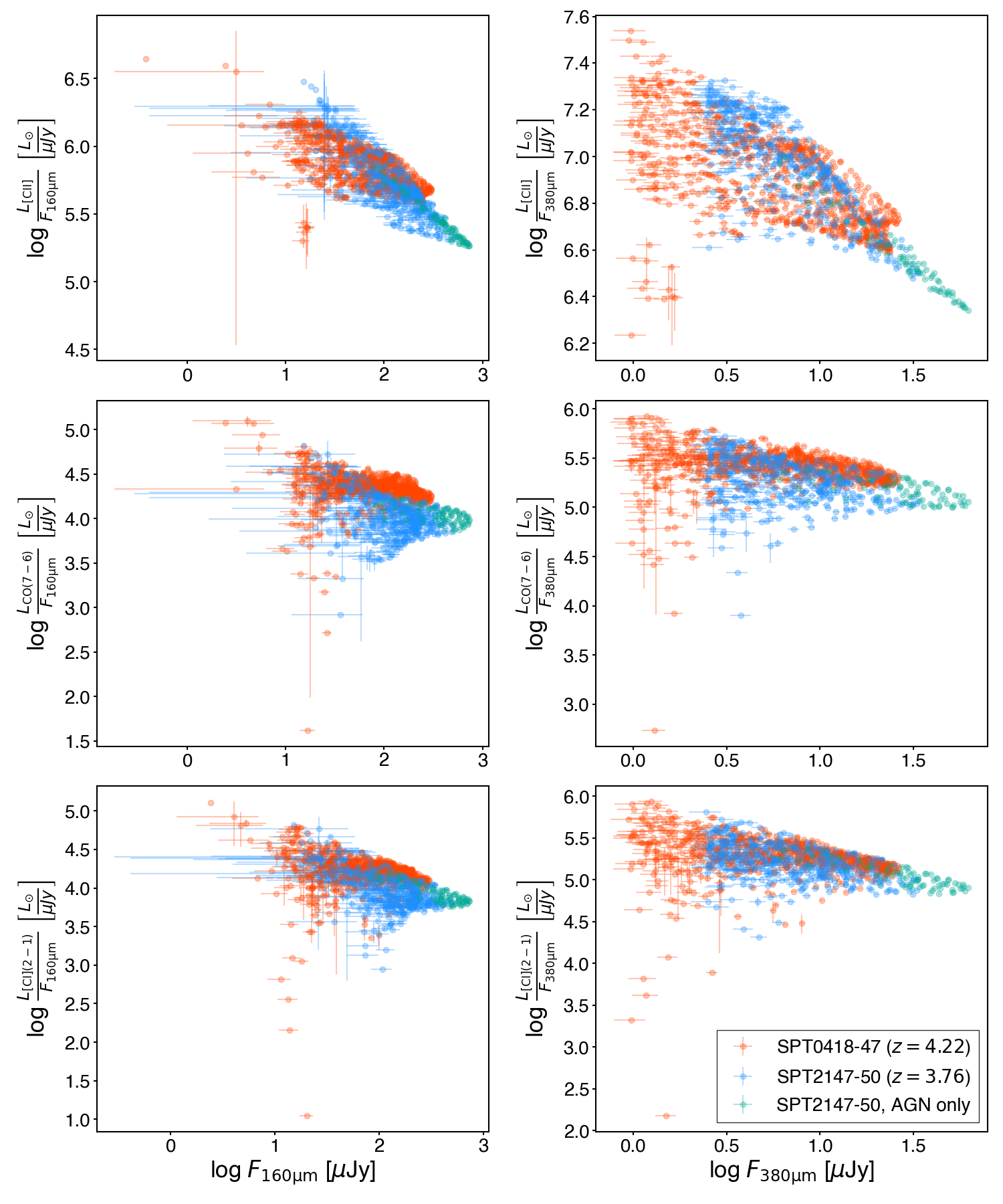}
    \caption{Line luminosity over continuum flux ratios at 160$\mu$m (left) and 380$\mu$m (right) for \cii (top), \coseven (middle), and \citwo (bottom).}
    \label{fig:SPT2147-50_flux_fractions}
\end{figure*}

Lastly, we investigated whether a signature of the AGN could be spotted by probing the dust (160, 380 $\mu$m continuum) against different line luminosities on spatially resolved scales. We show this comparison in Figure \ref{fig:SPT2147-50_flux_fractions}. Comparing line ratios in each source in both 160 $\mu$m and 380 $\mu$m flux per pixel does not reveal any stark differences between SPT0418-47 and SPT2147-50. The slope of $L_{\rm line} / F_{\rm cont}$ for both sources appears to be the same, with the AGN pixels located in the same region as the rest of the pixels in SPT2147-50. It is possible to proxy dust temperature by taking the flux ratio of two different continuum bands \citep[as in, for example,][]{spilker2023}, but a more technical comparison would be enabled by observing a third band of continuum with ALMA at $<$ 0.5\arc resolution, as fitting graybody functions to three bands of continuum imaging would output the spatially resolved dust temperature. We ultimately conclude that we cannot currently take advantage of our high-resolution ALMA observations, both in the source or image plane, to definitively determine the presence of an AGN in SPT2147-50, and this will require future work either with ALMA or by observing other windows of the electromagnetic spectrum via observatories like \textit{JWST}.

\end{appendices}

\end{document}